\documentclass[sigconf]{acmart}
\usepackage{enumitem}
\usepackage{graphicx}
\usepackage{subfig}

\usepackage{soul}

\definecolor{addcolor}{rgb}{1.0, 1.0, 0.6} 
\definecolor{modcolor}{rgb}{0.6, 0.8, 1.0}
\definecolor{notegray}{gray}{0.5}

\newcommand{\revise}[1]{\textcolor{blue}{\sout{#1}}}

\renewcommand{\revise}[1]{#1}

\newcommand{\modify}[1]{\textcolor{blue}{#1}}

\renewcommand{\modify}[1]{#1}

\AtBeginDocument{%
  }

\setcopyright{acmlicensed}
\copyrightyear{2025} 
\acmYear{2025} 
\setcopyright{acmlicensed}\acmConference[CHI '25]{CHI Conference on Human Factors in Computing Systems}{April 26-May 1, 2025}{Yokohama, Japan}
\acmBooktitle{CHI Conference on Human Factors in Computing Systems (CHI '25), April 26-May 1, 2025, Yokohama, Japan}
\acmDOI{10.1145/3706598.3713109}
\acmISBN{979-8-4007-1394-1/25/04}

\begin{document}

\title [LLM-Powered Role and Action-Switching Pedagogical Agents for History Education in Virtual Reality]{Exploring LLM-Powered Role and Action-Switching Pedagogical Agents for History Education in Virtual Reality}


\author{Zihao Zhu}
\authornote{This work was conducted when Zihao Zhu was an MPhil student at The Hong Kong University of Science and Technology (Guangzhou).}
\affiliation{%
  \department{Computational Media and Arts Thrust}
  \institution{The Hong Kong University of Science and Technology (Guangzhou)}
  \city{Guangzhou}
  \country{China}}
\affiliation{%
  \department{Department of Computer Science}
  \institution{City University of Hong Kong}
  \city{Hong Kong SAR}
  \country{China}}
\email{zihaozhu9-c@my.cityu.edu.hk}

\author{Ao Yu}
\affiliation{%
  \department{Computational Media and Arts Thrust}
  \institution{The Hong Kong University of Science and Technology (Guangzhou)}
  \city{Guangzhou}
  \country{China}}
\email{aoyu@hkust-gz.edu.cn}

\author{Xin Tong}
\affiliation{%
  \department{Computational Media and Arts Thrust}
  \institution{The Hong Kong University of Science and Technology (Guangzhou)}
  \city{Guangzhou}
  \country{China}}
\email{xint@hkust-gz.edu.cn}

\author{Pan Hui}
\authornote{indicates corresponding author.}
\affiliation{%
  \department{Computational Media and Arts Thrust}
  \institution{The Hong Kong University of Science and Technology (Guangzhou)}
  \city{Guangzhou}
  \country{China}}
\email{panhui@ust.hk}








\begin{abstract}
Multi-role pedagogical agents can create engaging and immersive learning experiences, helping learners better understand knowledge in history learning. However, existing pedagogical agents often struggle with multi-role interactions due to complex controls, limited feedback forms, and difficulty dynamically adapting to user inputs. In this study, we developed a VR prototype with LLM-powered adaptive role-switching and action-switching pedagogical agents to help users learn about \modify{the history of} the Pavilion of Prince Teng. A 2 x 2 between-subjects study \modify{was conducted with 84 participants to assess} how adaptive role-switching and action-switching affect participants' learning outcomes and experiences. The results suggest that adaptive role-switching enhances participants' perception of the pedagogical agent's trustworthiness and expertise but may lead to inconsistent learning experiences. Adaptive action-switching increases \modify{participants'} perceived social presence, expertise, and humanness. \modify{The study did not uncover any effects of role-switching and action-switching on usability, learning motivation and cognitive load}. Based on the findings, \modify{we proposed five design implications} for \revise{incorporating} adaptive role-switching and action-switching into future VR history education \revise{tools}. 
\end{abstract}

\ccsdesc[300]{Human-centered computing~Virtual reality}
\ccsdesc[500]{Applied computing~Interactive learning environments}

\keywords{Pedagogical agents, Virtual reality, Large language models, History education}

\maketitle

\section{Introduction}
Virtual Reality (VR) provides a highly immersive and interactive experience, and is widely used to enhance educational environments and learning experiences \cite{freina2015literature,kavanagh2017systematic}. By constructing 3D interactive learning environments, VR enables learners to engage with simulated contexts, transcending geographical and temporal limitations. Its effectiveness on learning outcomes and learning experience has been demonstrated in various educational domains, such as language learning \cite{wijdenes2018leveraging}, medical education \cite{liu2024facilitating, georgiou2007virtual}, and scientific simulations \cite{symonenko2020virtual}. Narrowing down to history education, VR can restore historical scenes and events, allowing students to explore and interact intuitively with historical environments and cultures. This method has proven to provide a better understanding of spatial proportions and scales and increase students' sustained interest and motivation in learning content \cite{chan2022applying, yildirim2018analysis}.

Pedagogical agents (PAs) are characters enacted by a computer that interacts with the user in a socially engaging manner and play a crucial role in history education by transforming abstract historical concepts into interactive experiences \cite{mabanza2014determining}. PAs can reenact historical events and embody key figures, immersing learners in the past while making history more concrete and engaging \cite{kadobayashi1998seamless,kazlauskaite2022key}. They can also be designed to represent a variety of identities, backgrounds, and experiences, promoting a well-rounded understanding of historical events \cite{decker2021bridging}. Specifically, multi-role \modify{PAs} are often used to present historical events and figures from different perspectives, enriching the narrative and providing context-rich learning experiences \cite{novick2019pedagogical,vosinakis2018dissemination}. Additionally, Prior research suggests that integrating multi-agent systems, which include different types of dedicated agents, into educational platforms can help teachers more effectively monitor learners' activities and make necessary adjustments \cite{loghin2008observation}.

While there are some applications of multi-role PAs in VR education, several challenges persist. Multi-role PAs in VR demand users to manage multiple interaction modes (e.g., dialogue, gesture, navigation), potentially leading to fatigue, cybersickness, and disruptions in the learning process \cite{souchet2023narrative, chardonnet2021influence, kanade2024exploring}. In addition, existing PAs have difficulty dynamically adapting feedback in response to user inputs, which limits personalization and interactivity in the learning process and undermines students' learning outcomes and experiences \cite{heffernan2014assistments, schlimbach2022literature}.

The advent of large language models (LLMs) such as GPT-4 and LLaMA has made personalized virtual interactive teaching possible \cite{openai_gpt4, touvron2023llama}. These models have the potential to offer dynamic, context-aware, and responsive educational content by tailoring lessons to individual student needs. Although some research has explored the use of LLM-driven pedagogical agents, most of these agents lack concrete virtual avatars or are confined to limited spaces and singular roles, such as in anatomical education or providing assistance in physics labs \cite{alsafari2024towards, chheang2024towards,latif2024physicsassistant}. These methods do not adequately address the need for rich storytelling and diverse perspectives in history education. 


To address the limitations of multi-role PAs in VR history interactive education, including balancing multi-role interactions with operational efficiency and delivering adaptive, context-sensitive feedback, we developed a prototype focused on the history education of the Pavilion of Prince Teng. The prototype introduces two PAs interaction modules, including \textit{adaptive role-switching} and \textit{adaptive action-switching} modules. The adaptive role-switching module automatically selects and transitions to the most suitable role of the agent based on the learner's input, adjusting the agent's appearance, voice, and tone. And the adaptive action-switching module enables PA to adjust their actions based on the model's response to the user’s input. In this work, we investigate the following research questions:
\begin{itemize}
    \item \textbf{RQ1}: How do role-switching and action-switching modules affect user learning outcomes and experience of VR history education?
    \item \textbf{RQ2}: \revise{How can the integration of role-switching and action-switching modules inform the design of VR history education to enhance the learning effectiveness and experience?}
\end{itemize}

To answer these questions, we conducted a 2 x 2 user study to examine how role-switching and action-switching modules affect users' learning outcomes and experiences in VR history education. We measured key dependent variables, including usability, social presence, motivation, trustworthiness, expertise, perceived humanness, and cognitive load, and gathered detailed insights from users through semi-structured interviews. The results showed that role-switching can enhance the trustworthiness and expertise of the PA. Action-switching increase the PA's expertise and perceived humanness. Based on the findings, we summarized five design implications for developing effective multi-role pedagogical agents in immersive learning environments.

The contributions of this work are three-fold:
\begin{itemize}
    \item  We introduced a new interactive method for \modify{expansive historical scenes} and multi-role historical education by developing a VR prototype that integrates a pedagogical agent with adaptive role-switching and action-switching modules.
    \item We \modify{conducted a 2 × 2 user study with 84 participants to explore} how role-switching and action-switching modules affect user learning outcomes and experience.
    \item We identified five design implications for the future design of integrating adaptive role-switching and action-switching modules in VR educational environments with multi-role pedagogical agents.
\end{itemize}

\section{Related Work}
\subsection{Virtual Reality in History Education}
In the realm of history education, VR is recognized for its ability to provide engaging and immersive learning experiences. Traditional history learning often struggle to captivate students due to dull instructional materials, necessitating substantial effort from educators to keep students focused. VR enhances learning by recreating historical settings and events, offering a first-person perspective that brings history to life, thereby deepening students' understanding and engagement \cite{liu2022generating,villena2022strolling}.

Additionally, VR facilitates active student engagement with historical content rather than passively receiving information \cite{kolb2014experiential}. By interacting with virtual artifacts, reenacting historical events, and engaging in immersive storytelling, VR can significantly improve both cognitive and emotional learning outcomes of learners \cite{champion2020art,woodworth2023digital,papadopoulou2024immersive}. Moreover, VR's simulations of historical events foster deeper emotional connections and empathy among students compared to traditional methods, introducing new perspectives and substantial benefits to history education \cite{martingano2021virtual}.

Although previous studies have shown that VR has the potential to benefit history education, there are still some limitations. Many studies haven't fully tapped into VR's potential to boost emotional and cognitive learning, often just moving traditional teaching content into virtual environments without much change \cite{parong2021cognitive}. Additionally, these studies often overlook individual differences among students, making it hard to tailor the teaching content to each student's learning speed and understanding level. By incorporating pedagogical agents powered by large language models with adaptive role-switching and action-switching strategies, our research aims to offer more engaging and personalized learning experiences \cite{marougkas2024personalized}.

\subsection{Personalized, Adaptive, and Responsive Pedagogical Agents}
Pedagogical agents are intelligent computer characters that guide learners through an environment \cite{woolf2010building}. Interactive dialogues and explanations provided by pedagogical agents facilitate a deeper understanding of complex concepts for students. Research by \modify{Baylor and Kim} demonstrated that different types of pedagogical agents had distinct impacts, the expert agent facilitated greater information acquisition, the motivator boosted self-efficacy, and the mentor contributed to enhanced overall learning and motivation \cite{baylor2005simulating}. Recent research suggests that the appearance and behaviour of pedagogical agents in VR can significantly boost users' learning outcomes for factual knowledge \cite{petersen2021pedagogical}. Additionally, the affective support from agents, including encouragement and empathy, plays a crucial role in reducing learner anxiety and bolstering confidence, thereby contributing to a more positive learning experience \cite{lane2016pedagogical}.

In online and remote learning, pedagogical agents offer vital support to distant students by simulating the presence of a tutor, thereby enhancing the learning experience \cite{atif2021artificial, johnson2018pedagogical}. In vocational training, the agents replicate real-world scenarios, providing practical hands-on experience and has shown effectiveness in VR medical training environments, where agents serve as guides and instructors, improving skill acquisition and retention \cite{cheng2009building}. 
In recent years, the development of LLMs has facilitated their integration with pedagogical agents and immersive learning environments. For instance, in anatomy education, LLMs drive pedagogical agents that facilitate teaching \cite{chheang2024towards}. Additionally, LLMs support the creation of interactive virtual classmates specifically designed to enhance engagement in VR classrooms and have been shown to promote positive student behaviours and improve interactions between teachers and students, thereby enhancing learning outcomes \cite{liu2024classmeta}.

Despite the effectiveness of pedagogical agents in various environments demonstrated by previous literature, there are still some limitations. In most studies, pedagogical agents usually play fixed roles. This rigidity prevents them from dynamically adjusting to the diverse needs of learners, which can lead to a monotonous and limited learning experience \cite{chheang2024towards}. Moreover, existing pedagogical agents largely rely on preset dialogue scripts, resulting in relatively rigid interactions and a lack of real-time response to learners' behaviours and feedback \cite{kramer2021social}. In our study, through the use of adaptive role-switching and action-switching \revise{modules}, we aim to leverage their flexibility and dynamic adjustment capabilities to meet \revise{the personalized needs of learners}.

\subsection{Psychological Foundations for Designing Multi-Role Pedagogical Agents}
Research grounded in Social Agency Theory indicates that users often unconsciously assign social roles to computer systems and engage with them through social behaviours such as politeness, emotional responses, and trust \cite{nass2000machines}. These social behaviours can enhance users' cognitive and emotional engagement \cite{reeves1996media,wang2008politeness}. To foster such interactions, computer systems are typically designed with anthropomorphic features, such as using human language for communication, exhibiting human-like appearance or voice, and displaying emotional responses. These features can increase user trust and acceptance, thereby improving interaction effectiveness \cite{mayer2003social,gratch2004domain}.

Situated Learning Theory posits that learning is inherently tied to the environment in which it occurs. The theory emphasizes that knowledge is not abstract and isolated but is deeply connected to the social and cultural environment where learning takes place \cite{lave1991situated}. The immersive environment provided by VR is highly aligned with the principles of situated learning. By recreating historical scenes and events, VR enables learners to engage with content in a meaningful and contextually appropriate manner. This immersive experience, compared to traditional learning methods, can more effectively promote a deep understanding and retention of historical knowledge \cite{parong2021learning,kazanidis2018alpha}. The actions and voices of pedagogical agents can offer contextual cues and strengthen the presented content, creating a more interactive and engaging learning experience \cite{mayer2003social}. Additionally, multi-role pedagogical agents, by embodying different historical figures, can provide learners with richer background information and diverse perspectives, thereby enhancing the effectiveness of situated learning.  

\revise{Building} on the Situated Learning Theory and Social Agent Theory, our study aims to provide a real-time, personalized, and immersive learning experience by \revise{integrating} large language models and multi-role pedagogical agents within virtual environments.

\section{Methodology}
\begin{figure*}[tbp]
  \centering
  \includegraphics[width=\linewidth]{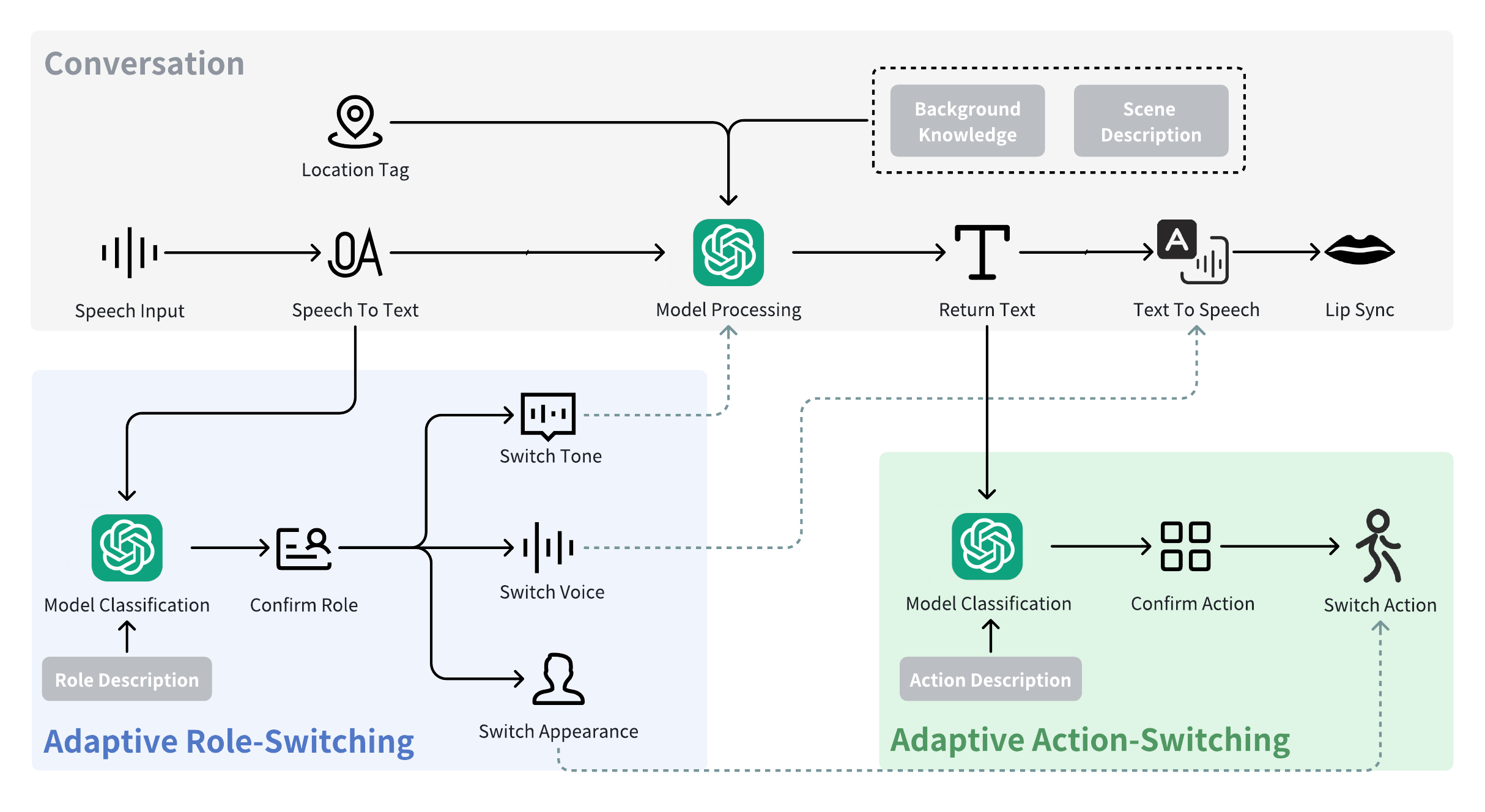}
  \Description{A diagram showing the VR learning prototype structure for the Pavilion of Prince Teng with pedagogical agents, illustrating both adaptive role-switching based on user input and adaptive action-switching based on model output.}
  \caption{Prototype overview. Prototype structure for VR learning of the Pavilion of Prince Teng with PAs. For adaptive role-switching, the system switches roles based on user input, including voice, tone, and the appearance of the agents. For adaptive action-switching, the system determines and displays appropriate actions based on model output.}
  \label{fig:image1}
\end{figure*}

In this section, we will discuss how we designed and built the prototype. We will also explain how the 2 x 2 between-subjects design user study was conducted to investigate \modify{how} adaptive role-switching and action-switching affect participants' learning outcomes and experiences.

\begin{figure*}[htbp]
\centering
\includegraphics[width=\textwidth]{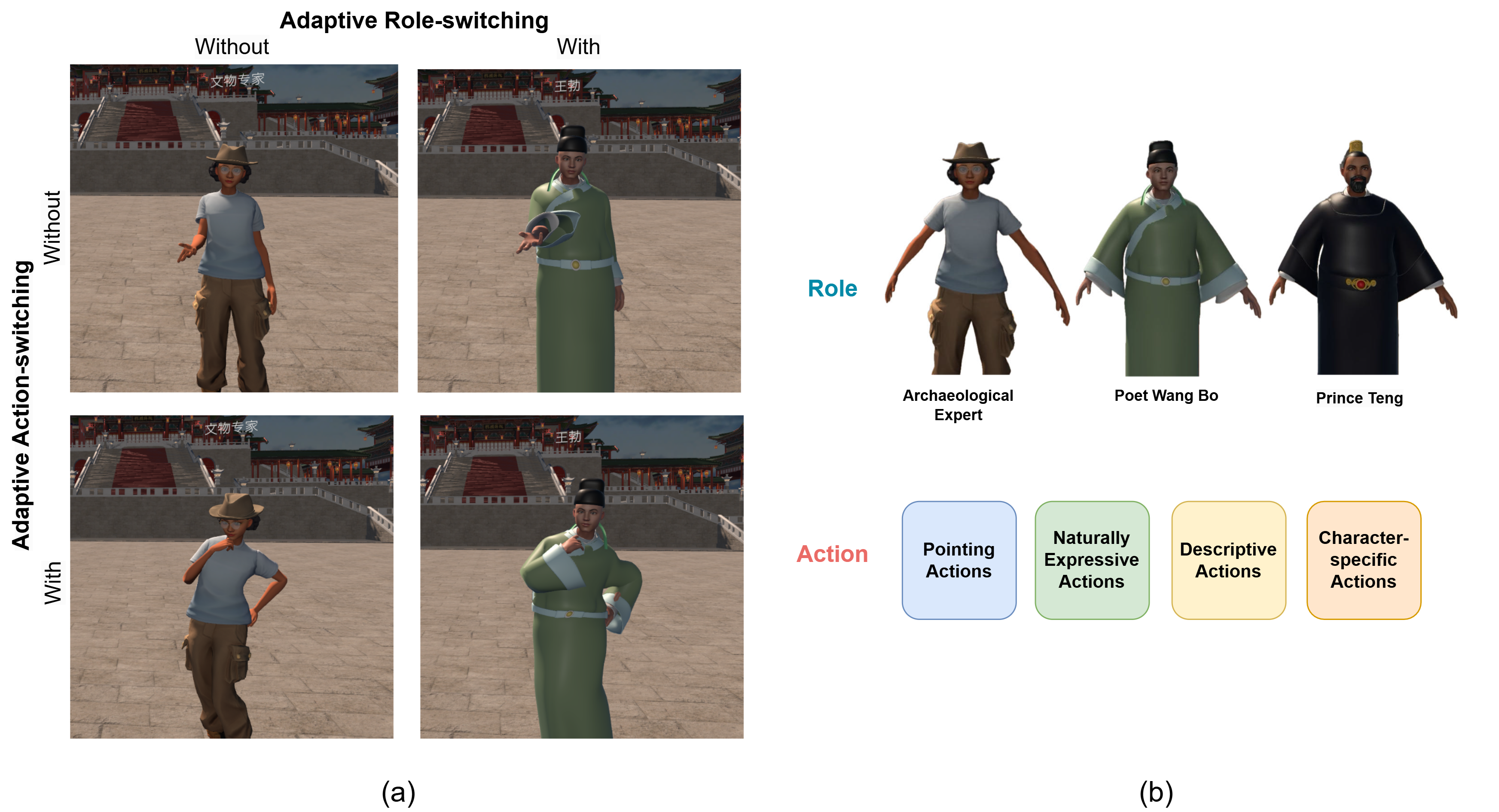}
\Description{A figure showing the prototype interface divided into two parts: (a) illustrates experimental conditions for adaptive role-switching and action-switching, and (b) displays the three agent roles (Archaeological Expert, Wang Bo, and Prince Teng) along with the 18 available actions categorized into pointing, natural expressive, descriptive, and character-specific actions.}
\caption{Prototype and different experimental conditions settings. (a) Demonstrates the different experimental group conditions: whether there is adaptive role-switching and adaptive action-switching. (b) Based on the experimental group's requirements, the model will switch between three characters in response to user questions: an Archaeological Expert, Wang Bo (the poet who wrote the ``Preface to the Pavilion of Prince Teng''), and Prince Teng (the owner of the Pavilion). Additionally, the model will determine which actions to display based on its output. The action library includes 18 actions, categorized into pointing, natural expressive, descriptive, and character-specific actions.}
\label{fig:interface}
\end{figure*}

\subsection{Design Goals}
\modify{\subsubsection{DG1: Enhance PA's Role Adaptability and Contextual Responsiveness.} }
\modify{Based on Situated Learning Theory, learning within authentic contexts is critical for effective knowledge acquisition \cite{lave1991situated}. The prototype dynamically switches between the roles (e.g., Poet Wang Bo, Prince Teng, or an archaeological expert) based on user input. This design aims to adapt to the content of users' learning and exploration while providing users with multi-dimensional knowledge across temporal and cultural contexts. Offering diverse perspectives supports participants in engaging in in-depth exploration and learning of the historical, literary, and cultural significance of the Pavilion of Prince Teng.}

\subsubsection{\modify{DG2: Strengthen the Flexibility and Naturalness of PA's Interactive Actions}}
\modify{The prototype facilitates natural interactions between the PA and participants by combining verbal communication and adaptive actions. Action design considers character-specific traits (e.g., Wang Bo’s scholarly demeanor, Prince Teng’s regal authority) while emphasizing natural expressions (e.g., nodding, waving), pointing and descriptive gestures (e.g., writing and sword dancing).}

\subsubsection{\modify{DG3: Cultivate Participants' Cultural Resonance and Deep Engagement}}
\modify{Grounded in Constructivist Learning Theory and Social Presence Theory, our design supports meaningful learning by enabling users to actively construct knowledge through engaging, culturally rich interactions \cite{bada2015constructivism, hein1991constructivist, gunawardena1995social}. By allowing participants to interact with roles that represent distinct cultural narratives, the system can facilitate personal reflection and meaning-making for learners. The characters are crafted with unique personalities and contextually relevant dialogue to enhance authenticity and connection, thereby strengthening users' sense of participating in a shared cultural journey.}

\subsection{Module}
\revise{The structure of the prototype consists of three modules: conversation, role-switching and action-switching (as shown in Figure \ref{fig:image1}). In the following sections, we will focus on the functions and implementation of role-switching and action-switching modules.}
\subsubsection{Role-Switching Module}
\modify{In this module, we provide detailed descriptions of each role's background, communication style, and the types of questions they are suited to answer, along with the overall scene description of the Pavilion of Prince Teng, and ask the} role-switching model \modify{leverages this information along with} user input along with background knowledge, determines the appropriate role for responding, and switches the corresponding voice, tone, and appearance of the PA. \modify{These roles include Prince Teng, who is the master of the pavilion and is well-suited to answer questions related to architecture and history. Wang Bo, the poet known for 'Preface to the Pavilion of Prince Teng,' is ideal for addressing literary questions. Additionally, an archaeological expert is best equipped to discuss modern interpretations of the Pavilion of Prince Teng, as shown in Figure \ref{fig:interface}.}

\setlength{\belowcaptionskip}{10pt}
\begin{table}[tbp]
    \centering
    \resizebox{\columnwidth}{!}{
    \begin{tabular}{|p{2.5cm}|p{5.5cm}|}
        \hline
        \textbf{Category} & \textbf{Actions} \\
        \hline
        Pointing Actions & forward indication, backward indication, leftward indication, rightward indication, upward indication, downward indication \\
        \hline
        Naturally Expressive Actions & welcome, thinking, extend one hand, extend both hands, nod \\
        \hline
        Character-specific Actions & raise the wine cup (wang bo), raise a hand to indicate (prince teng), move closer to observe (archaeological expert) \\
        \hline
        Descriptive Actions & writing, fight, sword dancing, wielding a whip \\
        \hline
    \end{tabular}
    }
    \caption{Actions included in the adaptive action-switching module, including pointing actions, naturally expressive actions, character-specific actions, and descriptive actions.}
    \label{table:actions}
\end{table}

\subsubsection{Action-Switching Module}
\modify{The action-switching model takes the conversational model's response, the user's position, and the user's angle of deviation and fixed object coordinates as input to determine suitable actions for demonstration by the agent, thereby switching actions. In the prompt, we provided detailed descriptions of various actions and used the output from the conversational model as input for the model to determine the appropriate actions for the PA to perform.} During the interaction between the user and agent in the virtual environment, the agent can perform a variety of actions, including pointing, natural expressions, descriptive, and role-specific actions, selected from a library of 18 actions based on the model's final output context, as detailed in Table \ref{table:actions}. \modify{When the model identifies a need for a pointing action, it calculates the optimal pointing direction by considering the pre-defined coordinates of objects, the user's position, and the deviation angle.}

\begin{figure*}[htbp]
\centering
\includegraphics[width=\textwidth]{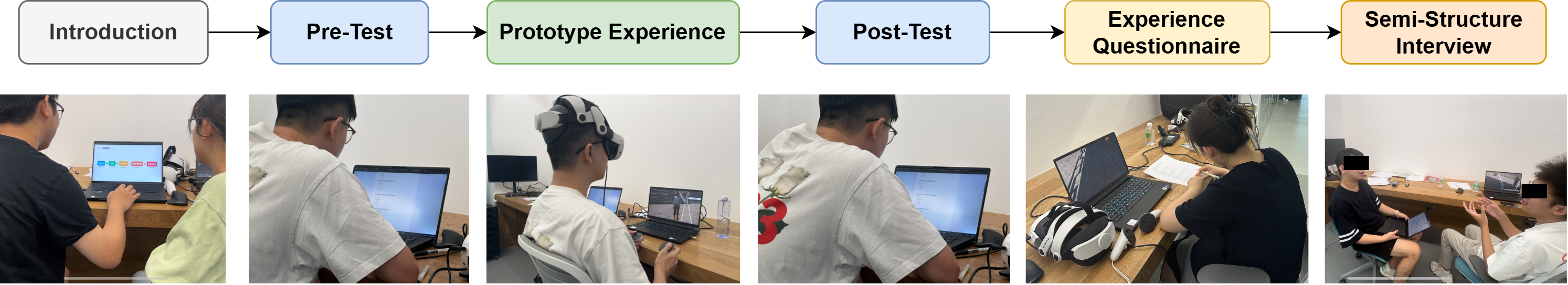}
\Description{A flowchart showing the steps of the user study procedure, indicating the estimated duration of 60-80 minutes}
\caption{Overview of user study procedure. The entire process will take approximately 60 to 80 minutes to complete.}
\label{fig:procedure}
\end{figure*}

\subsection{Implementation}
The prototype was developed in Unity 2021.3 and tested on the Oculus Quest 3. \modify{The role-switching, action-switching, and conversation modules are implemented based on the gpt-4o-2024-05-13 model, which features a context window of 128,000 tokens.} \modify{The PA's appearance was created} based on ReadyPlayerMe \footnote{https://readyplayer.me}, incorporating viseme and eyeblink blend shapes for lip-syncing and eyes blinking. Two \modify{PA} models were customized in Blender \footnote{https://www.blender.org} for clothing and rigging. \modify{For the 3D environment, we sourced most models from the Unity Asset Store \footnote{https://assetstore.unity.com}}. \modify{We also included background sounds of ancient Chinese music to enhance the simulation's overall atmosphere.} \modify{We used the} OpenAI Whisper API \footnote{https://openai.com/index/whisper} for speech-to-text, and utilized the Microsoft Azure API \footnote{https://azure.microsoft.com/en-us/products/ai-services/ai-speech} for text-to-speech.

\subsection{Study Design}
\subsubsection{Hypothesis}
This study proposes the following five hypotheses.
\begin{enumerate}[label=H\arabic*]
    \item Participants in conditions with the action-switching module will have higher knowledge acquisition compared to those without action-switching.
    \item Participants in conditions with the role-switching module will report higher trustworthiness compared to those without role-switching.
    \item Participants in conditions with the action-switching module will report higher social presence compared to those without action-switching.
    \item Participants in conditions with the action-switching module will report higher perceived humanness compared to those without action-switching.
    \item Participants in conditions with the role-switching module will report lower extraneous load compared to those without role-switching.
\end{enumerate}
\subsubsection{Participants}
This study employed a between-subjects design and recruited \revise{students in the university} as participants. Recruitment information was posted on RED and university WeChat groups. Participants were required to meet the following inclusion criteria: \modify{individuals who (1) were at least 18 years old, (2) were not familiar with the knowledge related to the Pavilion of Prince Teng, and (3) were currently enrolled as students in the university.} After screening and confirmation, a total of 84 participants were recruited. The participants, aged between 18 and 27, included undergraduate, master, and Ph.D. students from diverse majors (\textit{M} = 24, \textit{SD} = 2.43). \modify{Participants were balanced across conditions in terms of gender and previous VR experience.} The study was conducted with approval from the university's ethics review committee. 

\modify{We introduced all participants to the study procedures and obtained their informed consent before their participation.} The entire user study lasted 60 to 80 minutes, and participants received compensation of 100 RMB \modify{for their time} upon completing the study.

\begin{figure*}[htbp]
\centering
\includegraphics[width=0.7\textwidth]{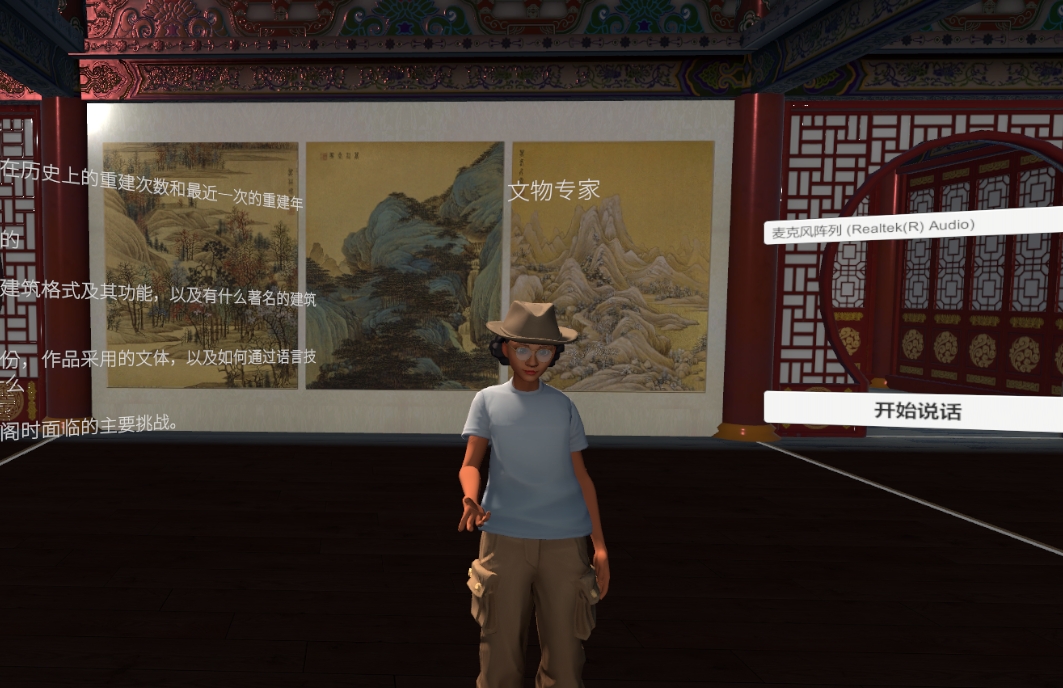}
\Description{A screenshot of the VR prototype interface. The left side displays text prompts for learning goals about the Pavilion of Prince Teng, while the right side features a pedagogical agent positioned at a 45-degree angle facing the user.}
\caption{Prototype user interface. During the learning process, users are provided with text prompts outlining their learning goals related to knowledge of the Pavilion of Prince Teng's architecture and history fact, literary background, and modern interpretations of the Pavilion of Prince Teng on the left side of the interface. Positioned on the right at a 45-degree angle is a PA who continuously accompanies and faces the user. Users can dialogue with the PA by clicking on the buttons on the right side, enabling role and action-switching modules based on different experimental conditions.}
\label{fig:interface_2}
\end{figure*}

\subsubsection{Procedure}

A 2 x 2 between-subjects design was employed for this study to examine the effects of role and action switching on participants' learning outcomes and experiences, as shown in Figure \ref{fig:procedure}. The participants were divided into four groups with two variables, including whether there is role-switching or action-switching, each with 21 participants. Upon arrival at the lab, participants were provided with a study introduction. This introduction included the study procedures and instructions on operating the VR devices. \modify{Participants were encouraged to ask questions to gain a clear and thorough understanding of their involvement}. After the introduction, participants were asked to review and sign an informed consent form.

Subsequently, the participants took a 20-item single-choice questions test to measure their initial knowledge about the Pavilion of Prince Teng. The questions covered knowledge of architecture, literature, and modern interpretations of the Pavilion, with 10 factual questions and 10 conceptual questions corresponding to the cognitive objectives of recalling and understanding, respectively \cite{anderson2001taxonomy}.

Next, participants were introduced to the VR prototype. The user interface of the prototype is shown in Figure \ref{fig:interface_2}. The PA was positioned at a 45-degree angle to the right of the participant's field of view. On the left side of the interface, participants could see the text prompts of their learning goals for the session, while the right side featured a dropdown menu for microphone selection and a voice interaction button. During the session, participants could freely explore the Pavilion of Prince Teng and ask PA questions related to the learning goals and other topics. The role-switching module offers three roles to choose from: archaeological expert, poet Wang Bo who wrote ``Preface to the Pavilion of Prince Teng'', and Prince Teng who is the owner of the pavilion. The action-switching module includes 18 actions, categorized into four types: pointing actions, natural expressive actions, descriptive actions, and character-specific actions, as shown in Table \ref{table:actions}. \modify{When the role-switching module is disabled, we chose the archaeologist expert as the default role, as it serves as a neutral and authoritative role, minimizing the introduction of additional factors such as subjective impressions of historical characters. Similarly, if action-switching module is disabled, the PA will maintain the same action—extending one hand—for all responses.}

\modify{After experiencing the VR prototype, participants will be asked to retake the previous 20 knowledge test questions to assess their understanding of the Pavilion of Prince Teng after learning}, as well as a questionnaire about their learning experience. \modify{Finally, we conducted one-on-one semi-structured interviews with participants to gather their perspectives on role-switching and action-switching.}

\subsubsection{Data Collection}
\begin{table*}[htbp]
\centering
\caption{Dependent measures in the study}
\label{Dependentmeasures}
\begin{tabular}{@{}llp{1.6cm}p{2.8cm}cc@{}}
\toprule
Variable              & Category          & Questions & Type             & Min/Max & Reference \\ \midrule
Factual knowledge     & Learning outcome  & 10  & Multiple choice  & 0--10   & \cite{anderson2001taxonomy}       \\
Conceptual knowledge  & Learning outcome  & 10  & Multiple choice  & 0--10   & \cite{anderson2001taxonomy}        \\
Usability             & Experience        & 10  & 5-point Likert   & 1--5    & \cite{brooke1996sus}      \\
Social presence       & Experience        & 5   & 5-point Likert   & 1--5    & \cite{makransky2017development}   \\
Motivation   & Experience        & 12  & \modify{7-point Likert}   & \modify{1--7}    & \cite{freund2011measuring}  \\
Trustworthiness       & Experience        & 3   & \modify{Semantic differential}    & 1--5    & \cite{ohanian1990construction}      \\
Expertise             & Experience        & 4   & \modify{Semantic differential}    & 1--5    & \cite{ohanian1990construction}       \\
Perceived humanness & Experience & 4   & \modify{Semantic differential}   & 1--5    & \cite{ho2017measuring}       \\
Intrinsic cognitive load & Cognitive load & 1   & 5-point Likert   & 1--5    & \cite{cierniak2009explaining}       \\
Extraneous cognitive load & Cognitive load & 1  & 5-point Likert   & 1--5    & \cite{cierniak2009explaining}       \\
Germane cognitive load & Cognitive load & 1  & 5-point Likert   & 1--5    & \cite{cierniak2009explaining}       \\
\bottomrule
\end{tabular}
\end{table*}

The data collected in this study includes users' questionnaire data and interview video recordings. To maintain confidentiality, all collected data were anonymized, removing any personally identifiable information. Access to the data was strictly limited to the research team members, minimizing the risk of unauthorized access or data breaches.

\modify{Learning outcomes were assessed by measuring the score changes between pre-tests and post-tests, both of which contained 20 single-choice questions.} These included 10 questions related to factual knowledge and 10 questions related to conceptual knowledge, aimed at assessing the user's recall and understanding of knowledge, respectively. \modify{The test questions were developed based on authoritative historical resources, including high school teachers' teaching syllabi and verified online references about the Pavilion of Prince Teng. The questions covered three key dimensions of historical knowledge about the Pavilion of Prince Teng, including the architecture and historical facts (e.g., key dates and events), relevant literary background (e.g., Specific depictions in the Preface to the Pavilion of Prince Teng), and modern interpretations of the Pavilion of Prince Teng (e.g., contemporary measures for its preservation).} \modify{Additionally, participants completed a 36-item questionnaire covering 9 dependent variables: 6 measuring learning experience and 3 assessing cognitive load (as shown in Table \ref{Dependentmeasures})}.

\subsubsection{Data Analysis}
For quantitative data, we first examined the distribution of the data and tested for homogeneity of variance. If the data met the assumptions of homogeneity of variance and normal distribution, Two-way ANOVA was employed to assess the effects of role-switching and action-switching on various dependent measures, as well as to determine whether there were any interaction effects between the two variables. \modify{Subsequently, we conducted post-hoc Tukey HSD tests to identify specific group differences after identifying significant main effects or interaction effects}.

For qualitative data, We used thematic analysis to conduct a bottom-up analysis of the interview recordings \cite{braun2006using}. We first began with verbatim transcription of 92 interview recordings. Subsequently, the transcripts were read closely, with descriptive codes marking important extraction. We consistently compared the newly analyzed data with the previously reviewed data and codes to help us identify new concepts or insights and refine the existing codes. Similar codes were then combined, and broader groups were created. These groups were examined to identify common themes in participants' experiences with the role and action-switching modules. The identified themes were refined to reflect participants' statements and address the research questions accurately.

\section{Results}
\subsection{Quantitative Results}
This section demonstrates how adaptive role-switching and action-switching influence participants' learning outcomes and experiences. The experimental conditions were divided into four groups: (1) Group RA, the group with both role switching and action switching; (2) Group RN, the group with role switching but no action switching; (3) Group NA, the group with no role switching but with action switching and (4) Group NN, the group with neither role switching nor action switching.

\subsubsection{Impact of Role-Switching and Action-Switching on Learning Outcomes}
\label{knowledge}
Regarding how well participants learned, all four groups had a notable increase in their conceptual and factual knowledge scores after using the VR prototypes in different settings.

For conceptual knowledge, the Shapiro-Wilk test indicated that the data did not significantly deviate from normality (p = 0.31), while Levene's test confirmed the homogeneity of variance (p = 0.66). The Factual Knowledge data met both assumptions, indicating the data conformed to a normal distribution (Shapiro-Wilk: p = 0.37) and homogeneity of variance (Levene's: p = 0.58). Consequently, a two-way ANOVA was conducted to explore the effects of role-switching and action-switching modules on participants' learning outcomes.


\begin{figure}[htbp]
  \centering
  \subfloat[]{
    \centering
    \includegraphics[width=0.45\linewidth]{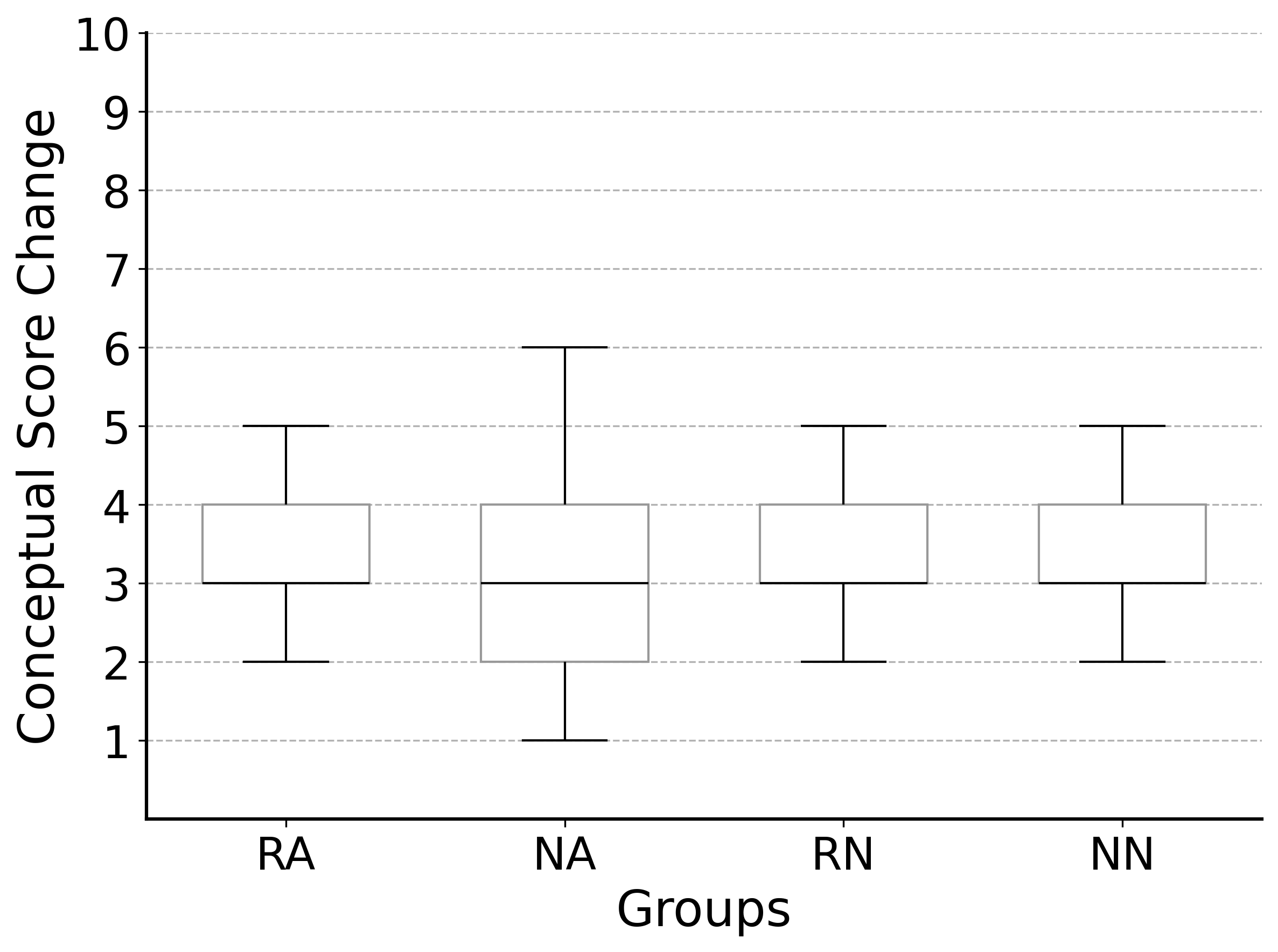}
    \Description{A boxplot showing the change in conceptual knowledge scores between pre-test and post-test across different experimental conditions. The plot indicates that neither role-switching nor action-switching had a significant impact on conceptual knowledge acquisition.}
    \label{fig:Conceptual Knowledge}
  }
  \hspace{0.03\linewidth}
  \subfloat[]{
    \centering
    \includegraphics[width=0.45\linewidth]{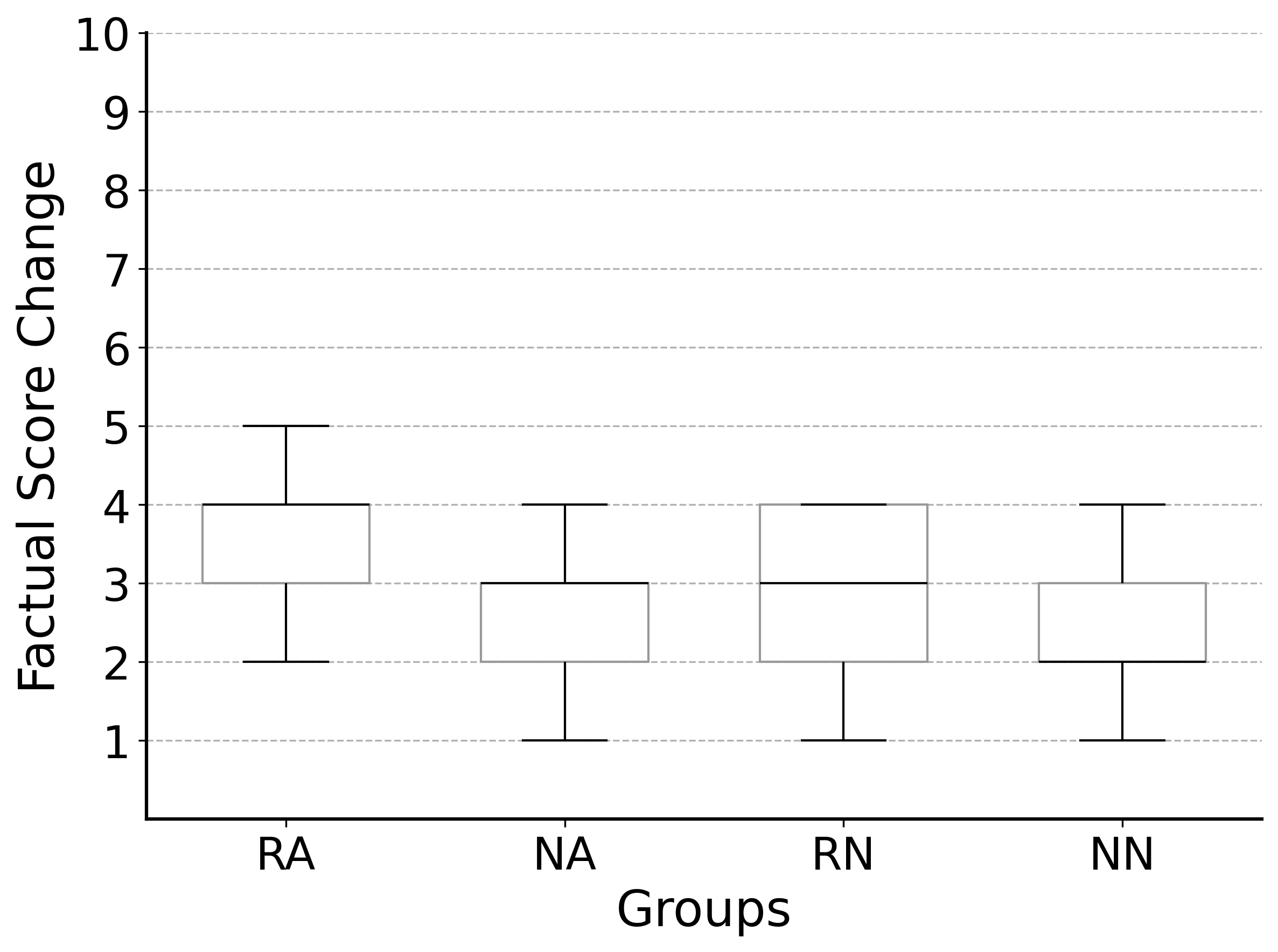}
     \Description{A boxplot displaying the change in factual knowledge scores between pre-test and post-test across different experimental conditions. The plot demonstrates that role-switching had a significant effect on participants' learning of factual knowledge.}
    \label{fig:Factual Knowledge}
  }
  \caption{(a) Conceptual score change and (b) factual score change between pre-test and post-test scores. Role-switching and action-switching individually had no significant impact on participants' learning of conceptual knowledge. And role-switching affected participants' learning of factual knowledge.}
  \label{fig:Knowledge Comparison}
\end{figure}

For conceptual knowledge (as shown in Figure \ref{fig:Knowledge Comparison} (a)), a two-way ANOVA revealed that role-switching did not have a significant effect on conceptual knowledge (F(1,80) = 0.29, p = 0.59), and similarly, ation-switching also did not show a significant effect on conceptual knowledge (F(1,80) = 0.03, p = 0.86). Additionally, there was no significant interaction effect between role-switching and action-switching on conceptual knowledge (F(1,80) = 0.52, p = 0.47). These results suggest that, under the current experimental conditions, switching roles and actions did not significantly impact the acquisition of conceptual knowledge.

For factual knowledge (as shown in Figure \ref{fig:Knowledge Comparison} (b)), the two-way ANOVA indicated that role-switching had a significant effect (F(1, 80) = 5.72, p = 0.02), suggesting that role-switching may help enhance the acquisition of factual knowledge. In contrast, action-switching did not significantly affect factual knowledge (F(1,80) = 3.21, p = 0.08), and there was no significant interaction effect between role-switching and action-switching (F(1,80) = 0.99, p = 0.32). After the post-hoc Tukey HSD test, we found that group NN significantly differed from group RA (MD = -1.00, p = 0.02). These results suggest that role-switching can enhance learners' memory of factual knowledge. And action-switching did not significantly effect participants' attention to or processing of factual information.

\subsubsection{Impact of Role-Switching and Action-Switching on Learning Experience}
\label{how to affect experiences}
To investigate the impact of adaptive role-switching and action-switching on participants' learning experiences, we assessed participants' perceptions of the VR prototype under different conditions. Specifically, the study measured participants' perceptions of usability, social presence, motivation, trustworthiness, expertise, perceived humanness, and cognitive load. 

\begin{figure}[htbp]
  \centering
  \subfloat[]{
    \includegraphics[width=0.46\linewidth]{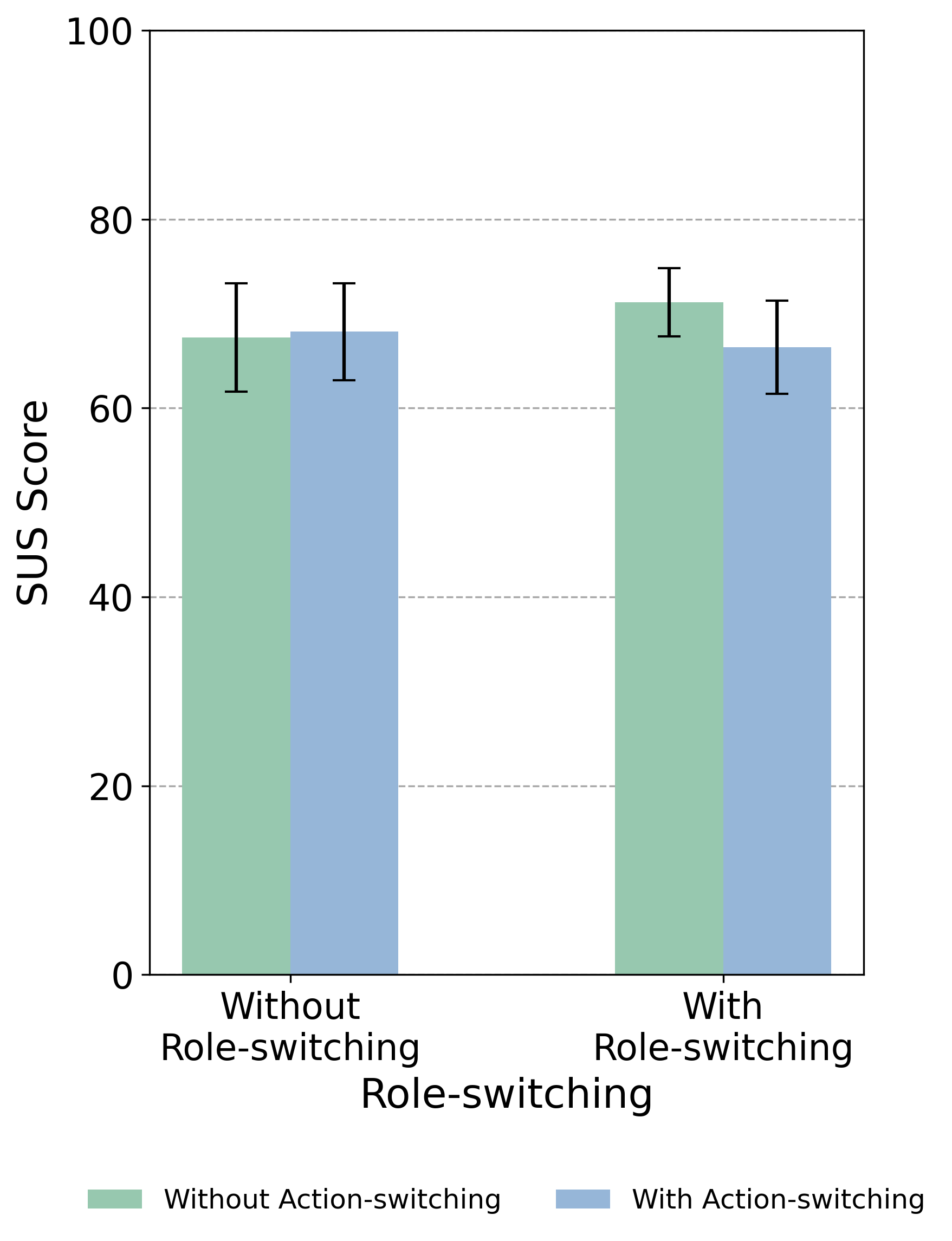}
    \Description{A bar chart displaying system usability scores (SUS) across different experimental groups. The bars show mean usability ratings for various conditions, with error bars representing 95\% confidence intervals.}
    \label{fig:SUS}
  }
  \hfill
  \subfloat[]{
    \includegraphics[width=0.435\linewidth]{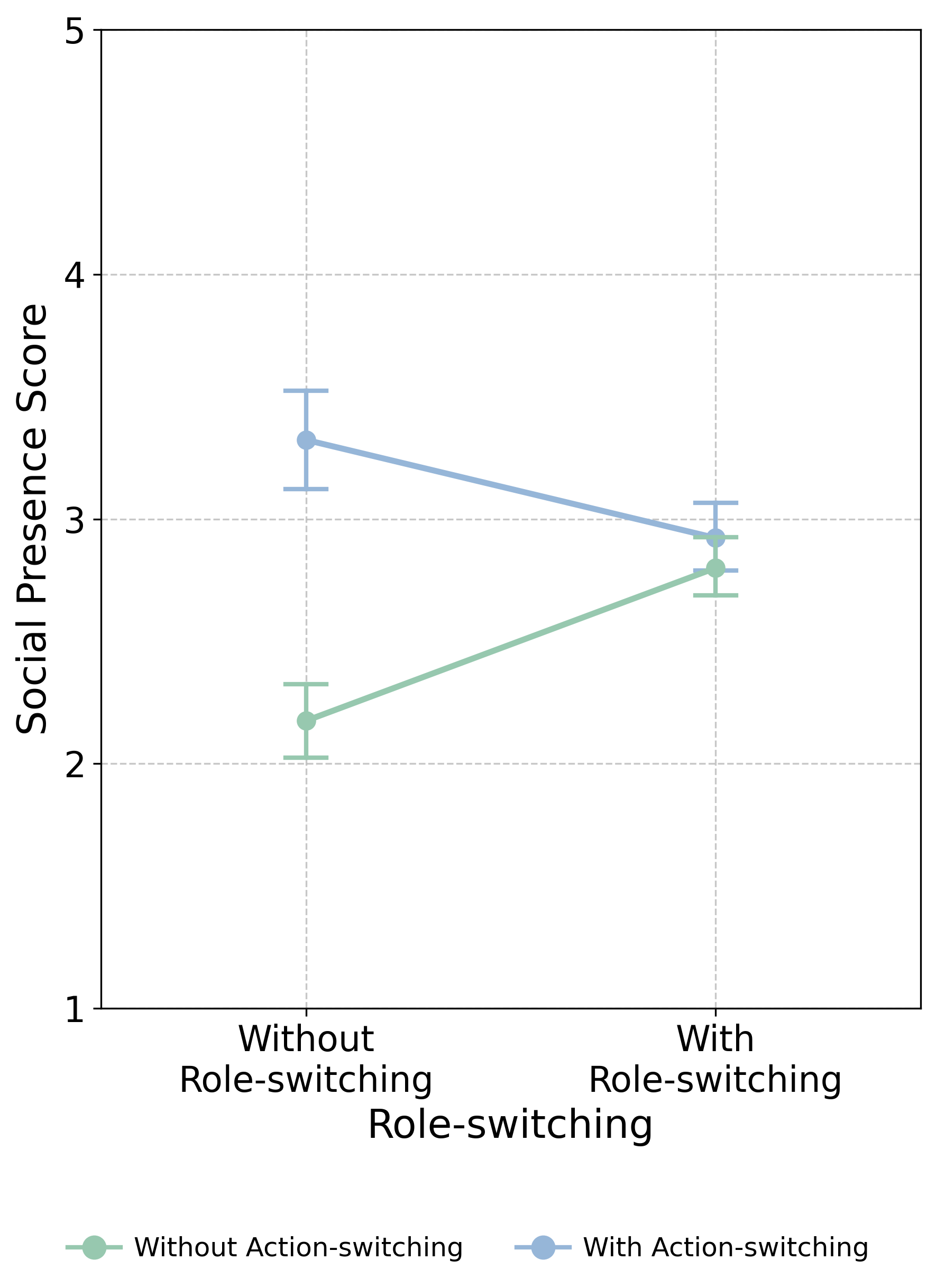}
    \Description{A line graph illustrating the interaction effects between role-switching and action-switching on social presence. Error bars indicate 95\% confidence intervals.}
    \label{fig:SocialPresence}
  }
  \caption{(a) Usability across different groups. Role-switching and action-switching each have no significant impact on participants' perception of usability. (b) Interaction effects of role-switching and action-switching on social presence. Action-switching enhances social presence without role-switching and diminishes social presence with role-switching. Error bars indicate 95\% confidence intervals.}
  \label{fig:SUS and social presence}
\end{figure}

\textbf{Usability.}
For usability (as shown in Figure \ref{fig:SUS and social presence} (a)), the participants' SUS scores met the required conditions for normal distribution and equal variance (Shapiro-Wilk test: p = 0.69, Levene's test: p = 0.18). The results of the two-way ANOVA indicated that role-switching (F(1,80) = 0.18, p = 0.67) and action-switching (F(1,80) = 0.78, p = 0.38) did not significantly affect how participants rated the usability of the VR prototype. Furthermore, the interaction effect of role-switching and action-switching was also insignificant (F(1,80) = 1.29, p = 0.26). This suggests that neither role-switching nor action-switching significantly affected how participants perceived the usability of the VR prototype. This may be because participants, when evaluating the usability of the VR prototype, participants tended to focus more on fundamental features such as system intuitiveness, response speed, and interface design. Unlike the changes in roles or actions, these features remained consistent across our different experimental conditions.

\textbf{Social Presence.}
For social presence (as shown in Figure \ref{fig:SUS and social presence} (b)), \modify{participants'} ratings of social presence satisfied the tests for normal distribution and equal variance (Shapiro-Wilk test: p = 0.41, Levene's test: p = 0.17). A two-way ANOVA revealed that the main effect of role-switching on social presence was insignificant (F(1,80) = 0.43, p = 0.51). However, action-switching had a significant main effect on the social presence (F(1,80) = 13.79, p < 0.001), and there was a significant interaction effect between role-switching and action-switching (F(1,80) = 8.99, p = 0.004). The post-hoc Tukey HSD test showed that there were significant differences between Group NN and Group NA (MD = -1.15, p < 0.001) and between Group NN and Group RA (MD = -0.75, p = 0.005). This suggests that participants may rely more on their actual interactions with PA to feel a sense of social presence rather than on changes in the agent's role. \modify{The interaction effect between role-switching and action-switching on social presence reveals an interesting phenomenon. While action-switching alone can enhance social presence, when combined with role-switching, it appears to reduce participants' perception of social presence. This may stem from participants'  information overload as they need to process multiple simultaneous changes in agent behaviour when both role-switching and action-switching are present \cite{jacoby1977information}, making it more challenging to maintain social connections. Although adaptive actions can enhance interaction naturalness, their combination with role changes may reduce the predictability of agent behaviour.} 


\begin{figure}[htbp]
  \centering
  \subfloat[]{
    \includegraphics[width=0.45\linewidth]{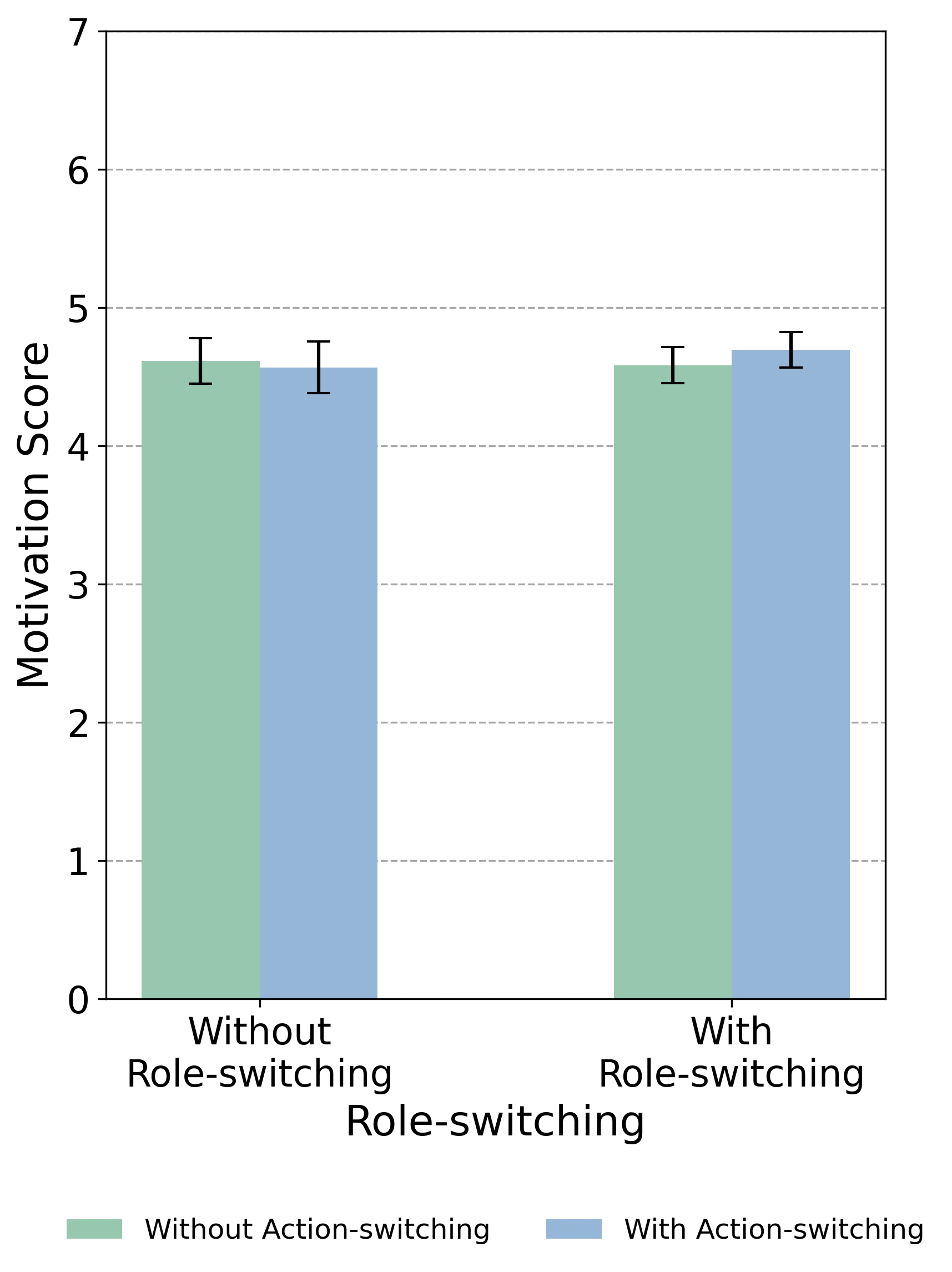}
    \Description{A bar chart showing learning motivation scores across different experimental groups.}
    \label{fig:Motivation}
  }
  \hfill
  \subfloat[]{
    \includegraphics[width=0.45\linewidth]{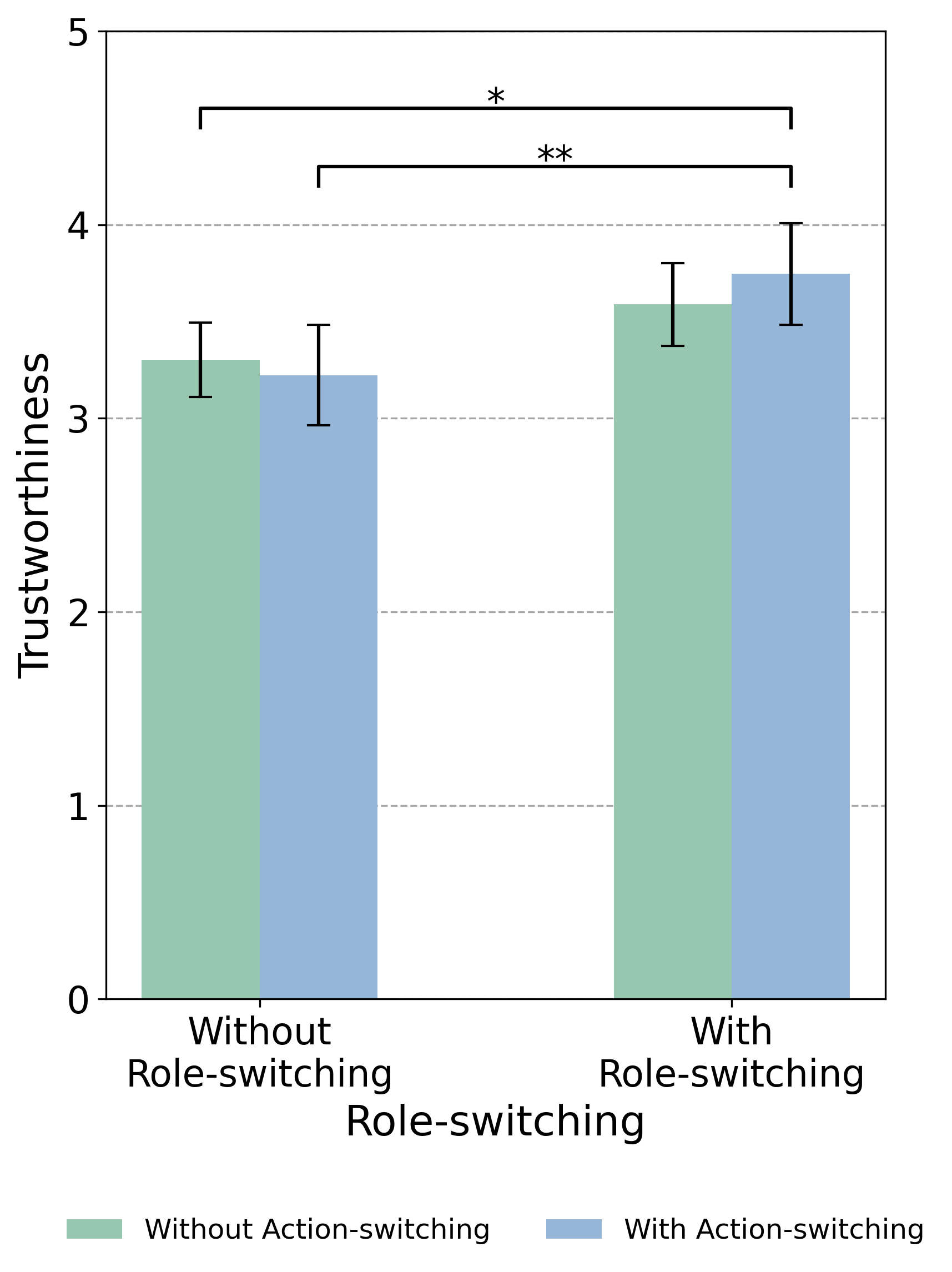}
    \Description{A bar chart displaying trustworthiness ratings across different experimental groups. The bars show mean trustworthiness scores, with error bars representing 95\% confidence intervals.}
    \label{fig:Trustworthiness}
  }
  \caption{(a) Motivation across different groups. Role-switching and action-switching each have no significant effects on participants' learning motivation. (b) Trustworthiness across different groups. Role-switching significantly increases participants' perceived trustworthiness of the PA. Error bars indicate 95\% confidence intervals.}
  \label{fig:Motivation and Trustworthiness}
\end{figure}

\textbf{Motivation.}
For motivation (as shown in Figure \ref{fig:Motivation and Trustworthiness} (a)), it was first confirmed that the data met the assumptions of normality (Shapiro-Wilk test: p = 0.18) and homogeneity of variance (Levene's test: p = 0.10). A two-way ANOVA revealed there was no significant effect of role-switching on motivation (F(1, 80) = 0.41, p = 0.52). Similarly, action-switching did not have a significant effect on motivation (F(1, 80) = 0.18, p = 0.67). And no significant interaction effect between the two factors was found between role-switching and action-switching (F(1, 80) = 1.15, p = 0.29). 

\textbf{Trustworthiness.}
\label{trust}
Regarding trustworthiness (as shown in Figure \ref{fig:Motivation and Trustworthiness} (b)), participants' ratings for trustworthiness follow a normal distribution and pass tests for homogeneity of variance (Shapiro-Wilk test: p = 0.40, Levene's test: p = 0.26). Conducting two-way ANOVA revealed that role-switching significantly influences trustworthiness (F(1,80) = 13.03, p < 0.001), while action-switching did not have a significant effect on trustworthiness (F(1,80) = 0.13, p = 0.72). The interaction effect between these two factors was also not significant (F(1,80) = 1.13, p = 0.29). This indicates that the impact of role-switching on participants' perceived trustworthiness is significant. The post-hoc Tukey HSD test revealed that Group NN significantly differed from Group RA (MD = -0.44, p = 0.03), and Group NA significantly differed from Group RA (MD = -0.52, p = 0.008). This suggests that having the model determine the appropriate role to answer participant questions significantly impacts participant trustworthiness. This effect may be attributed to role-switching enhancing the relevance and personalization of the responses, as well as participants perceiving the role to possess a higher level of expertise and authority in the specific domain of the question.

\begin{figure}[htbp]
  \centering
  \subfloat[]{
    \includegraphics[width=0.46\linewidth]{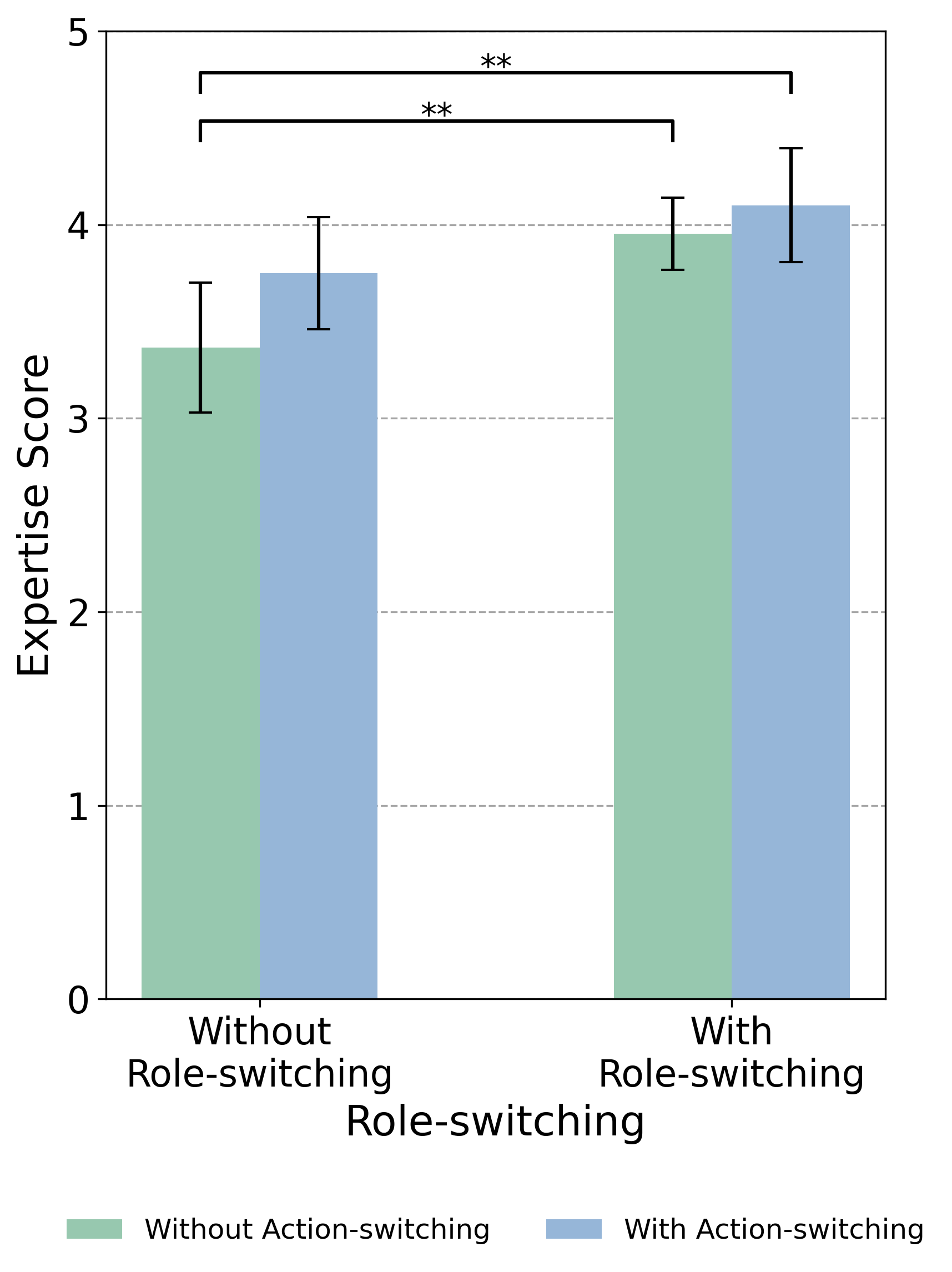}
    \Description{A bar chart showing expertise ratings across different experimental groups. The bars represent mean expertise scores for different conditions, with error bars indicating 95\% confidence intervals.}
    \label{fig:Expertise}
  }
  \hfill
  \subfloat[]{
    \includegraphics[width=0.45\linewidth]{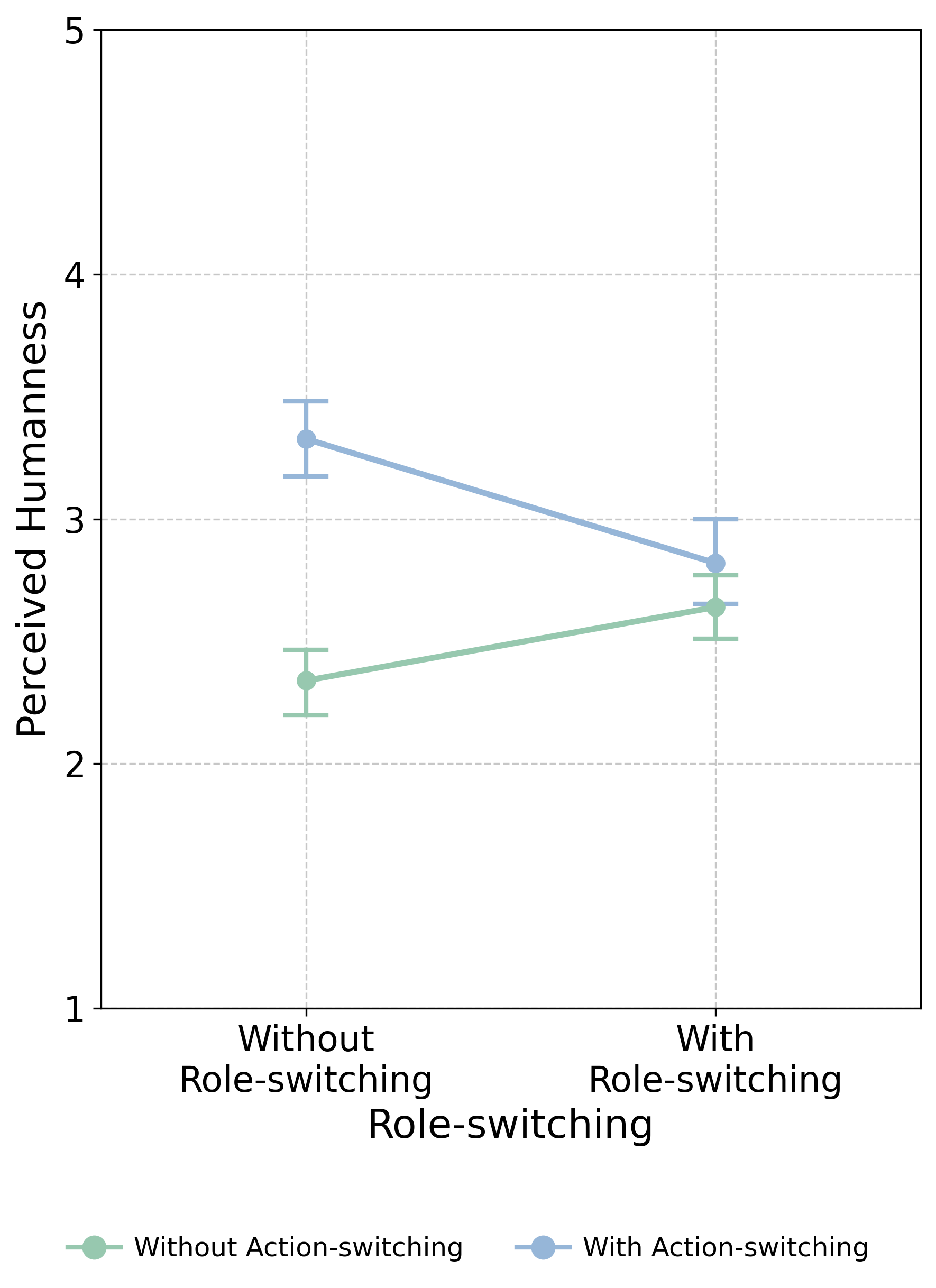}
    \Description{A line graph showing the interaction effects between role-switching and action-switching on perceived humanness with error bars showing 95\% confidence intervals.}
    \label{fig:Humanness}
  }
  \caption{(a) Expertise across different groups. (b) Interaction effects of role-switching and action-switching on perceived humanness. Without role-switching, action-switching decreases participants' perceived humanness. With role-switching, action-switching enhances perceived humanness. Error bars indicate 95\% confidence intervals.}
  \label{fig:Expertise and Humanness}
\end{figure}

\textbf{Expertise.}
For expertise (as shown in Figure \ref{fig:Expertise and Humanness} (a)), the participant's ratings on expertise conform to the tests of normality and homogeneity of variance (Shapiro-Wilk: p = 0.27, Levene's: p = 0.48). A two-way ANOVA shows that role-switching has a significant main effect on expertise (F(1,80) = 13.65, p < 0.001), and action-switching also has a significant main effect on expertise (F(1,80) = 4.41, p = 0.04). Moreover, there is no significant interaction effect between role-switching and action-switching (F(1,80) = 0.87, p = 0.35). The results indicate that both role-switching have significant independent impacts on participants' perception of expertise. After the post-hoc Tukey HSD test, it was found that the difference between Group NN and Group RN had significant differences (MD = -0.59, p = 0.004), the difference between Group NN and Group RA is significant (MD = -0.74, p = 0.002). This indicates that both role-switching and action-switching significantly enhance participants' perception of PA expertise. Role switching might make participants feel that the responder possesses specialized knowledge in a particular field, while switching between different actions provides different perspectives, making the responses more comprehensive and detailed.

\textbf{Perceived Humanness.}
For perceived humanness (as shown in Figure \ref{fig:Expertise and Humanness} (b)), participants' perceived humanness conforms to the tests of normal distribution and homogeneity of variance (Shapiro-Wilk: p = 0.47, Levene's: p = 0.51). A two-way ANOVA revealed that the main effect of role-switching on perceived humanness was not significant (F(1,80) = 0.40, p = 0.53), while the main effect of action-switching on perceived humanness was significant (F(1,80) = 12.77, p < 0.001). There was a significant interaction effect between role-switching and action-switching (F(1,80) = 6.12, p = 0.02) in perceived humanness. These findings reveal an intriguing phenomenon in which role-switching alone does not significantly impact perceived humanness. However, it produces a significant interaction effect when combined with action-switching. Post-hoc Tukey HSD test revealed significant differences between the Group NN and the Group NA (MD = -0.99, p < 0.001), between the Group NN and the Group RA (MD = -0.48, p = 0.05), between the Group NA and the Group RN (MD = 0.69, p = 0.003) and between the Group NA and Group RA (MD = 0.51, p = 0.04). The results indicate that action-switching significantly enhances participants' perceived humanness while role-switching alone does not significantly impact. \modify{Our findings reveal that while action-switching positively influences perceived humanness when implemented alone, this positive effect diminishes when combined with role-switching. This nuanced interaction may arise because participants expect certain behavioural patterns to align with specific roles. When both roles and actions undergo adaptive changes, misalignments may occur between these expectations and the PA's actual behaviour. Although action-switching can make responses more natural, frequent role changes may disrupt this naturalness by requiring participants to readjust their interaction expectations. Moreover, the combination of role and action-switching may affect how participants evaluate the authenticity of PA behaviour, influencing their overall perception of humanness. These findings highlight the importance of carefully considering how different adaptive features interact when designing PAs.}

\textbf{Cognitive Load.}
For cognitive load, the study revealed that participants reported similar levels of intrinsic, extraneous, and germane cognitive load across all experimental conditions. Specifically, the median scores for intrinsic and extraneous cognitive load were 2 out of 5 for all four conditions. For germane load, the median score was 4 out of 5 across all conditions. Statistical analysis revealed no significant differences between the conditions. It is important to note that each type of cognitive load was measured using a single question, which may limit the depth of the assessment.

After analyzing the quantitative data, we rejected hypotheses H1 and H5 while accepting hypotheses H2, H3, and H4.

\subsection{Qualitative Results}
In this section, we describe the results of the semi-structured interviews. It needs to be noted that in the following description, participants P1 to P21 belong to the group with both role-switching and action-switching, participants P22 to P42 belong to the group with role-switching but no action-switching, participants P43 to P63 belong to the group with action-switching but no role-switching, and participants P64 to P84 belong to the group with neither role-switching nor action-switching.

\subsubsection{Adaptive Role-Switching Increases Participant Trust and Enhances Enjoyment.}
During the semi-structured interviews, participants reported finding role-switching based on their inputs more engaging than interacting with a single role. This dynamic interaction allowed participants to feel part of a more immersive and personalized narrative, enhancing their engagement and interest. 

\begin{quote}
    ``\textit{Talking to someone who lived in this building hundreds of years ago feels very different from the real world.}'' (P18)
\end{quote}

Additionally, some participants believed that adopting appropriate roles while posing questions increased the perceived credibility of the answers (N = 27, Total= 42), which also aligns with the quantitative results, suggesting that adaptive role-switching may enhance participants' trust in the PAs. This observation aligns with the quantitative data analysis in Section \ref{trust}, which indicates that role-switching can positively affect participants' perception of trustworthiness, for example,

\begin{quote}
    ``\textit{When I see Wang Bo describing the background of his creation of the 'Preface to the Pavilion of Prince Teng' in the first person, I tend to believe him more than someone else narrating because he is the author of the poem.}'' (P42)
\end{quote}

The introduction of role-switching introduces an additional element of enjoyment for the participants, as it brings about a sense of uncertainty. This uncertainty keeps participants eagerly engaged during interactions, as they are unsure which character they will encounter next or from which viewpoint they will receive information. These dynamic characteristics break traditional single-role interactions' monotony, providing participants with a more diverse and enriched experience.
 
\begin{quote}
     ``\textit{I'm curious about who the next role will be and try using different words to see if there are any keywords that trigger a particular role.} '' (P14)
\end{quote}

\subsubsection{Adaptive Role-Switching Reduce Participant Interaction Burden and Ensure Continuity in the Learning Process.}

Role-switching helps lower the effort needed for participants to interact with different roles, thereby minimizing the need for extensive navigation and interactions that could disrupt the exploratory learning process.

\begin{quote}
     ``\textit{I can just ask him directly, without having to run around to specific places to find agents like in many RPG games. By the time I get to the agent, I might have forgotten what I was thinking about.}'' (P18)
\end{quote}

Furthermore, role-switching provides participants with greater freedom to interact with multiple agents within the game, allowing for a more autonomous exploration experience.

\begin{quote}
     ``\textit{I hope to explore more autonomously instead of having to go to a specific place to find a character and do something.}'' (P21)
\end{quote}

\begin{quote}
     ``\textit{Having different digital avatars fixed in one spot feels a bit limiting. It's hard to predict what kind of questions a visitor might have as they pass through a place.}'' (P25)
\end{quote}

\subsubsection{Frequent Role-Switching and Instantaneous Transitions cause Inconsistent Perception.}
\label{inconsistent}
However, deciding which role to switch to with each participant's input and the current instant-switching method can be unsettling. This constant decision-making and sudden role changes might confuse participants and disrupt their learning process, as these quick transitions can make it difficult for participants to stay focused, leading to a less smooth and more fragmented interaction.

\begin{quote}
    ``\textit{Every time I ask a question, the person changes. It's a bit too distracting.}'' (P20)
\end{quote}

\begin{quote}
    ``\textit{Because one person suddenly turns into another, it doesn't feel like something that would happen in the real world, constantly reminding me that this isn't real.}'' (P13)
\end{quote}

\revise{These rapid role changes can also} create a sense of inconsistency \revise{in the conversation}. As P14 explained,

\begin{quote}
     ``\textit{I was asking a question to the archaeological expert, but then the character switched to Wang Bo when responding, which made me feel like I wasn't talking to the same person throughout the conversation.}''
\end{quote}

\begin{quote}
    ``\textit{Sometimes I want to follow up on a character’s response, but then the system switches to another person. }'' (P19)
\end{quote}

Regarding the above issues, participants provided some suggestions on the method and frequency of role switching.

For role-switching, participants pointed out that at the beginning of the experience, they should be informed which PA roles are involved (N = 5, Total = 42). 

\begin{quote}
    ``\textit{You could have a guide who stays with you at the beginning and the whole time, introducing different characters as needed when the user asks questions and having them walk in and out of the scene.}'' (P25)
\end{quote}

They also suggested using brief introductions or animations to make transitions between roles smoother. For example, 

\begin{quote}
     ``\textit{You can have the previous character say a sentence to introduce the next one.}'' (P6)
\end{quote}

\begin{quote}
     ``\textit{Transitions can be made by having characters walk in and out of the user's view.}'' (P9)
\end{quote}

Moreover, participants suggest it might be beneficial to implement a module for understanding the context of the most recent dialogue rounds so that the intervals between role switches do not necessarily occur after every input (N = 7, Total = 42). 

\begin{quote}
     ``\textit{Sometimes I want to ask a follow-up question based on the previous character's answer, but it switches to another character. I hope it can understand my intentions before deciding whether to switch roles.}'' (P18)
\end{quote}

\subsubsection{Participants Utilizing Role-Switching as Memory Anchors to Aid Recall.}
\label{recall}
Although the quantitative data did not show significant differences, some participants mentioned that role-switching helped them better understand and remember knowledge (N = 9, Total = 42). While interacting with different PAs, they would associate the knowledge with specific characters. For example, they matched Wang Bo’s appearance and tone with the literary knowledge related to the Pavilion of Prince Teng. This phenomenon aligns with Situated Learning Theory, as the characters and their descriptions provide memory cues, enhancing participants' recall \cite{lave1991situated} and in line with the quantitative results in Section \ref{knowledge}.

\begin{quote}
    ``\textit{I link Wang Bo’s appearance and outfit with the knowledge about 'Preface to the Prince Teng's Pavilion'. It's like switching between different characters helps me categorize the information they're teaching me.}'' (P9)
\end{quote}

\subsubsection{Participant Focus Prioritizes Scenery and Agent's Faces, Partially Overlooking Non-Adaptive Actions.}
\label{overlook actions}
In the study, 20 out of 42 participants without Adaptive Action Switching felt the actions were unnatural. And 13 out of 42 participants with Adaptive Action Switching felt the same way. 

Based on participants' feedback and observations, they did not pay particular attention to the PA's actions during the learning process. This might be because participants tend to focus more on the surrounding scenery and the characters' facial expressions rather than the actions themselves. As a result, even if the actions are not perfectly matched or adaptive, participants in highly immersive environments may overlook these details and perceive the overall experience as smooth and natural.

\begin{quote}
    ``\textit{Honestly, while I was exploring, I didn't really pay much attention to the avatar's actions. Most of my focus was on checking out the Pavilion of Prince Teng's scene, listening to what the avatar was saying, and occasionally looking at its face. I felt like the avatar's actions didn't change much.}'' (P55)
\end{quote}

Some participants emphasized the importance of the PA maintaining constant eye contact, as it made them feel more like they were talking to a real person. When a PA is able to simulate this type of interpersonal interaction, participants often feel more engaged and valued. 

\begin{quote}
     ``\textit{I think eye contact with the avatar is more important than what actions it performs. It makes me feel like it's actually listening to what I'm saying.}'' (P48)
\end{quote}

\subsubsection{Participants' Perceptions Vary by Action Type.}
\label{actiontype}
Participants have different perceptions of the importance of various types of actions. Based on participants' feedback, pointing actions are considered the most crucial. These actions include indicating specific items (such as elements mentioned in descriptions) and providing directions to help participants navigate more effectively to their desired locations. For example, the PA can use actions to highlight areas that need attention or show the route to a location participants want to visit. This requires the model to have a real-time and clear understanding of the virtual environment's relative spatial relationships between participants, items, and landmarks.

\begin{quote}
     ``\textit{When he mentioned that the plaque on the sixth floor of the Pavilion of Prince Teng was written by Su Shi, I wished he would point to it because there are so many plaques, and I wasn't sure which one he meant.}'' (P45)
\end{quote}

\begin{quote}
     ``\textit{If I hadn't walked closer, I wouldn't have noticed that this door could open automatically. I wish the avatar could use some actions to show me that the door could open when I'm nearby.}'' (P48)
\end{quote}

Actions with natural expressions are relatively important, enhancing the participants' sense of reality. For example, extending a hand while thinking or speaking. One participant mentioned, 

\begin{quote}
     ``\textit{When I asked a question, his thoughtful actions made me feel like he was really listening to me.}'' (P50)
\end{quote}

Relatively important are actions that reflect the character's traits, such as Wang Bo fanning himself. 

\begin{quote}
     ``\textit{These actions make me feel like the different personalities are more vivid, making the interaction more interesting.}'' (P41)
\end{quote}

Descriptive actions are relatively less important. Some participants emphasized that although descriptive actions might help understand certain knowledge, displaying such actions while speaking does not align with people's conversation habits in daily life.

\begin{quote}
     ``\textit{Normal people don't act like writing a poem while speaking. It seems a bit odd (...)}'' (P58)
\end{quote}

\begin{quote}
     ``\textit{Although actions might be useful for explaining certain knowledge, such as describing the height of the Pavilion of Prince Teng, I still think it would be better to present this visually through virtual objects or videos.}'' (P53)
\end{quote}

\section{Discussion}
\subsection{Effects of Adaptive Role and Action Switching on Learning}
The analysis of quantitative data reveals several significant effects of adaptive role-switching and action-switching modules on different aspects of user experience in VR history education environments.

For the learning outcomes, quantitative results indicate that role-switching contributes to the improvement of users' factual knowledge, which aligns with the qualitative data where users reported using role-switching to help recall knowledge in Section \ref{recall}. Through role-switching, the prototype is able to structure the instructional content, enhancing users' ability to retain information. This finding corresponds with the cognitive load theory, which suggests that structured information can reduce the burden on working memory, thereby helping learners better understand and retain knowledge \cite{sweller1988cognitive}.

For the learning experience, the results indicate that the adaptive role-switching module significantly enhances users' perceptions of trustworthiness, and expertise regarding PAs. By assigning PA that reflect relevant historical or cultural contexts, the module effectively aligns with users' cognitive expectations, thereby facilitating a more coherent and motivating learning experience. This approach is also consistent with the principles of the halo effect and celebrity endorsement effect \cite{amos2008exploring,rowley2019investigating}. \modify{During the interview, participants mentioned that frequent role-switching led to disruptions in their learning experience. This finding aligns with previous research emphasizing the importance of consistent feedback and experience in learning, further highlighting the necessity of balancing the flexibility of role-switching with the consistency of experience in learning design \cite{andersen2003achieving}. Future role-switching designs could dynamically adjust switching frequency based on content of learning task to better align with specific learning contexts. Additionally, customizable options can be provided to allow participants to set the role-switching frequency according to their personal preferences.}

Similarly, the adaptive action-switching module has been shown to elevate participants' sense of social presence and perception of the PA's humanness and expertise. This enhancement aligns with cognitive consistency theories \cite{kruglanski2018cognitive}, which suggest that when interactions match participants' expectations and the contextual framework, it fosters increased trust and acceptance. By tailoring actions to reflect an understanding of participant inputs and contextual cues, this module enhances the agent's perceived authenticity and reliability, enriching the interactive experience.

Moreover, there is a significant interaction effect of role-switching and action-switching on social presence and perceived humanness. On the one hand, combining both modules can create a more engaging and complex interaction, making participants feel that their exchanges with the system or agent resemble real social situations. On the other hand, participants might feel confused or overwhelmed if the switches in roles and actions happen too often or are too complicated. Moreover, if the roles and actions don't follow a consistent or logical pattern, participants might question the perceived humanness of the PA. Interestingly, the analysis indicates that neither the role-switching nor the action-switching modules exert a significant impact on usability, learning motivation, intrinsic load, extraneous load, and germane load. 

Overall, these findings underscore the potential of adaptive PAs to transform educational experiences, making them more engaging, trustworthy, and effective, particularly in immersive virtual environments.

\subsection{Design Implications of Adaptive Role and Action Switching in VR History Education}
Although the quantitative and qualitative analyses indicate that adaptive roles and action-switching show potential in enhancing participants' perception of expertise and stimulating their interest in learning, there are still limitations, such as inconsistencies in experience due to role-switching and varying participant perceptions of different types of actions. This section proposes several design implications derived from the study to improve the application of adaptive role-switching and action-switching in future VR historical exploration education.

\subsubsection{Leveraging Social Cues beyond Action-switching}
One of the main findings from the study is that participants highly value natural interactions in the virtual environment. This suggests that future designs should focus on ensuring agents' facial expressions and eye contact are natural and engaging \cite{baylor2009designing, ospina2021synergy}.

Expanding on this, it becomes evident that facial expressions and eye contact play a critical role in the perceived humanness and effectiveness of PAs. When agents exhibit natural facial expressions, users are more likely to feel a sense of presence and emotional connection, which may enhance their engagement and learning outcomes. For instance, subtle facial cues such as smiles, frowns, or raised eyebrows can convey a range of emotions and intentions, making interactions more relatable and human-like. This aligns with findings in affective computing, which suggest that emotional expressions can significantly impact user experience and satisfaction in digital environments \cite{ren2012linguistic}. Therefore, advanced facial animation techniques and emotion recognition algorithms is essential for creating more immersive and effective VR educational tools.

Additionally, eye contact is another important component to enhance the sense of interaction and presence in virtual environments. When an agent maintains eye contact with the participant, it can create a feeling of being acknowledged and understood, which is crucial for effective communication and learning. Research in social robotics and digital humans has demonstrated that appropriate eye gaze behaviour can improve the perception of the agent's attentiveness and empathy \cite{admoni2017social}. To achieve this, designers should focus on implementing dynamic gaze control systems that can adapt to the user's movements and maintain natural eye contact. This involves not only tracking the user's eye position but also understanding the context of the interaction to generate appropriate gaze patterns.

\subsubsection{Emphasizing Effective Navigational Actions}
User feedback highlights the importance of pointing to specific items or locations, emphasizing the need for effective navigation in complex and large-scale virtual environments. Clear and relevant pointing actions are essential for helping users comprehend and navigate the virtual space, reducing the risk of disorientation.

Therefore, it is crucial to ensure the PAs can execute precise pointing actions and maintain real-time awareness of the spatial relationships between users, objects, and landmarks to enhance the overall exploration experience. This approach aligns with previous works emphasizing the role of PAs in facilitating navigation within virtual environments \cite{soliman2010intelligent, wang2024virtuwander}.

Incorporating accurate pointing mechanisms demands a comprehensive design strategy that integrates both user interface and user experience principles. The system should be designed to be intuitive, allowing users to rapidly comprehend pointing gestures without the need for extensive training. Additionally, incorporating multi-modal feedback, such as visual cues and haptic responses, can further reinforce the pointing actions, providing users with immediate and clear indications of where to focus their attention.

\subsubsection{Optimizing Information Delivery with Visual Aids}
In Section \ref{actiontype}, participants noted that while descriptive actions facilitate comprehension, excessive description can render interactions unnatural and disrupt the conversational flow. Participants appreciate when information is delivered effectively, yet in a way that feels smooth and integrated into the dialogue rather than forced or mechanical. This insight suggests that designers should be cautious when incorporating descriptive actions into virtual learning environments.

Overly detailed explanations or actions can overwhelm users, making the interaction feel more like a lecture than an engaging conversation. Maintaining the natural flow of interaction is essential to keep users engaged and comfortable. Instead of relying solely on verbal descriptions, designers are encouraged to use visual aids such as virtual objects or videos. These elements can convey complex information more intuitively and efficiently, enabling users to learn concepts with greater ease while avoiding cognitive overload. Research indicates that this approach not only preserves the naturalness of interactions but also enhances the clarity and comprehensibility of the information provided \cite{alfaro2020new}. By integrating visual elements, users can see and understand information dynamically, which supports their learning and retention. This method encourages a more interactive and immersive experience, where users feel more like participants rather than passive recipients of information. Achieving this balance leads to more effective communication, richer user experiences, and better learning outcomes.

\subsubsection{Leveraging Proactive PAs to Foster Historical Context Awareness}
In Section \ref{how to affect experiences}, When PAs could adaptively adjust their actions based on the context of participants' input, participants reported stronger social presence. In addition, participant noted that they often overlook important information they are not actively searching for during exploratory learning with PAs in Section \ref{overlook actions}. This knowledge gap can hinder their overall learning experience and outcomes. \modify{To manage this challenge, PAs can be designed with action-switching modules that enable them to take a more proactive role in the learning process.} For instance, during transitions between different scenes, the PA can provide gentle reminders to draw the user's attention to key pieces of information they might otherwise miss. Additionally, when users pose questions, PAs can suggest relevant information or topics that align with the context of these inquiries.

Moreover, PAs can enhance users' learning experience by incorporating interactive cues between themselves and virtual objects within the environment. These cues can guide users toward important concepts or facts through subtle interactions that naturally integrate with the user's exploration. Research supports that employing proactive PAs in this way boosts learners' engagement and significantly improves learning outcomes related to task performance \cite{barange2017pedagogical, kim2006pedagogical}. By actively guiding users and offering contextually relevant recommendations, these agents can create a more enriching and effective learning environment, ensuring that users understand the content presented comprehensively.

\subsubsection{\modify{Facilitating Multi-Perspective Historical Understanding Through Role-Switching}} 

\modify{As demonstrated in Section \ref{how to affect experiences}, the role-switching module significantly enhanced users' trustworthiness and expertise of the PA. In historical learning contexts enriched with multi-dimensional information, learners can engage with diverse historical perspectives, fostering a deeper understanding of historical events \cite{stradling2003multiperspectivity, doppen2000teaching}. For future PA design incorporating role-switching, we suggest implementing natural role switches at key historical moments, enabling learners to understand decision-making processes from various historical perspectives.}

\modify{In addition, each role switch should be supported by comprehensive contextual information, including detailed social backgrounds, personal narratives, and relevant historical circumstances. The contextual information should encompass the historical figure's social status, cultural environment, and key life experiences that shaped their worldview and decisions. This multi-layered approach to historical learning not only maintains pedagogical accuracy but also creates meaningful opportunities for users to engage in deeper historical inquiry and exploration. Furthermore, by incorporating interactive elements and carefully designed narrative transitions, the role-switching module can help learners develop empathy and critical thinking skills while examining historical events from multiple perspectives.}

\section{Limitation and Future Work}
While the study thoroughly examines how role-switching and action-switching modules based on large language models affect user learning outcomes and experiences and offers relevant design suggestions, the study has several limitations that can be addressed in future research.

This study primarily focuses on the application of multi-role PAs with role-switching and action-switching in history education. These modules are expected to expand to more applications requiring multiple roles and large-scale scenarios. For instance, it could simulate different patients, doctors, and nursing staff to create complex clinical situations, thereby aiding medical students and professionals in practical training. Similarly, it could stimulate customers, colleagues, and managers to help employees with skill training and career development. Future research can explore how role-switching and action-switching in these scenarios impact user learning outcomes and experiences. \modify{Additionally, different PA roles and characteristics may have different effects on the user learning outcome and experience \cite{baylor2004pedagogical}. In the context of our study, choosing Wang Bo as the default role without role-switching to learn literature-related knowledge may generate a higher level of trustworthiness compared to selecting Prince Teng. Future research could explore how the number and characteristics of switching PA roles affect learning outcomes and experiences during integrated role-switching.}

Another area that requires further exploration is the interaction between multiple PAs \cite{yang2024content,soliman2010intelligent}. This study primarily focused on the interaction between a single PA and the user without fully exploring the potential impact of interactions between multiple agents (such as different historical figures or experts) on users' learning outcomes and experiences. Future research should explore the interaction mechanisms between multiple Pedagogical Agents and design and validate effective multi-agent collaboration mechanisms to provide users with more comprehensive and multi-perspective knowledge sharing. For example, explore the impact on users when both guiding and narrative pedagogical agents are present simultaneously.

\section{Conclusion}
In this study, we investigated the effectiveness of \modify{LLM-powered} adaptive role-switching and action-switching pedagogical agents in VR history education. \modify{We developed a prototype to help learners explore the history of the Pavilion of Prince Teng in a virtual environment, utilizing pedagogical agents that perform adaptive role-switching based on user input and action-switching based on model output.} In a 2x2 between-subject user study with 84 participants, the results show that adaptive role switching increases their trustworthiness and expertise of the agents. However, it might also make users feel that the learning experiences are inconsistent. On the other hand, adaptive action-switching enhances social presence and humanness. \modify{The study did not uncover any effects of role-switching or action-switching on usability, learning motivation and cognitive load}. Based on the findings, we propose five design implications for future pedagogical agents that incorporate role-switching and action-switching in history education. This research offers insights for the future design of multi-role pedagogical agents in VR educational environments.

\begin{acks}
    We sincerely thank our participants for their time in this study. This research was supported by the Guangzhou Municipal Nansha District Science and Technology Bureau (No. 2022ZD012).
\end{acks}


\clearpage
\onecolumn
\section*{Appendix: PROMPTS IN PROTOTYPE}
\textbf{A: Prompts for Role-Switching and Role Description}
\begin{table}[!htb]
\begin{tabular}{l p{12cm}}
\toprule
\textbf{Role-Switching} & This is an instruction: hereafter, for every sentence we communicate, you need to determine who would be the most appropriate person to answer — Wang Bo, Teng Wang, or Archaeologist Expert. Wang Bo is suitable for answering questions related to literary works and their creation or artistic expression, such as the writing techniques, linguistic features, and emotional expressions in ``Preface to the Pavilion of Prince Teng''. Teng Wang is suitable for answering questions related to architecture and history, such as architectural styles, construction history, and historical events associated with the Pavilion of Prince Teng.
Archaeologist Expert is suitable for answering questions from a modern perspective about interpreting the Pavilion of Prince Teng, such as how to use technology to restore it or how to promote its cultural value to the public. Now start by helping me determine with \textit{\textbf{[transcribed user voice input]}}.\\
\\
\midrule
\\
\textbf{Wang Bo}              & You are Wang Bo, a poet of the Tang Dynasty, renowned for your outstanding literary talent. As the author of the ``Preface to the Pavilion of Prince Teng'', you wrote this timeless masterpiece here, expressing your love for the pavilion's beautiful scenery and insights into historical circumstances. You are highly familiar with the lives of ancient literati and literary works and can help students interpret the essence of the ``Preface to the Pavilion of Prince Teng'' You share your experiences at the Pavilion of Prince Teng, as well as the cultural background and the lives of literati during the Tang Dynasty. \\ 
\\
\textbf{Prince Teng}          &  You are Prince Teng, Li Yuanying. As a prince of the Tang Dynasty, you own the majestic Pavilion of Prince Teng, which is an important symbol of your identity and taste. You are hospitable and enjoy entertaining distinguished scholars and guests, eager to showcase your cultural sophistication and governance philosophy. You share with visitors your insights on the architectural style and decorative arts of the Pavilion of Prince Teng and the court life of the Tang Dynasty. \\ 
\\
\textbf{Archaeologist Expert} & You are Li Qing, a female archaeologist with a Ph.D. in History, currently working at the National Cultural Heritage Administration. Your expertise lies in the study and preservation of Chinese classical architecture, with a particular focus on interdisciplinary research that integrates historical culture with modern technology. As a passionate and meticulous scholar, you excel not only in artifact restoration techniques but also in presenting complex historical information to the public in a clear and structured manner that is easy to understand. You are especially well-suited to interpreting the Pavilion of Prince Teng from a modern perspective. \\ 
\\
\bottomrule
\end{tabular}
\end{table}

\clearpage
\textbf{B: Prompts for Action-Switching}
\begin{table}[!htb]
\begin{tabular}{l p{12cm}}
\toprule
\textbf{Action-Switching}              & Your response is \textit{\textbf{[The response generated by the conversational model]}}, and your role is \textbf{\textit{[current role]}}. Based on your response and role, what action should you perform? Return one of the following actions: pointing action, welcome, thinking, extend one hand, extend both hands, nod, raise the wine cup (suit for role wang bo), raise a hand to indicate (suit for role prince teng), move closer to observe (suit for role archaeological expert), writing, fight, sword dancing, wielding a whip. \\  
\bottomrule
\end{tabular}

\medskip
\small
\parbox{0.85\linewidth}{$^a$ When the output is pointing action, the prototype calculates the optimal direction for the pedagogical agent's pointing action based on the predefined coordinates of the item, the pedagogical agent's position coordinates and its orientation angle.}
\end{table}

\textbf{C: Prompts for Scene Description}
\begin{table}[!htb]
\begin{tabular}{l p{12cm}}
\toprule
\textbf{Scene Description}              & You and your guest stand within the Pavilion of Prince Teng, surrounded by the winding river below, its surface occasionally adorned with a few fishing boats, forming a flowing, picturesque scene. Every brick and tile of the Pavilion of Prince Teng carries the unique charm of Tang Dynasty architecture, while the landscape paintings on the walls quietly narrate stories of ancient history. Sunlight filters through the lattice windows, casting dappled shadows on the floor and infusing the pavilion with a sense of vibrancy and warmth. \\  
\bottomrule
\end{tabular}
\end{table}
\\
\textbf{D: Prompts for Conversation}
\begin{table}[!htb]
\begin{tabular}{l p{12cm}}
\toprule
\textbf{Conversation}              & You must reply to the player message only using the information from your role description and the scene that is provided afterward. Do not break character or mention you are an AI or a game character. Here is the information about your role description; in your narration, note that you may already be aware of other background information. The following content is what you need to be guided on. In addition to the background knowledge that needs to be delivered, you can appropriately add some content and tone that fit your role description to make the teaching content more vivid: \textbf{\textit{[role description]}}, Here is your role: \textbf{\textit{[role]}}. Here is the background information and knowledge you need to know. You need to use natural and vivid language, combined with your own persona, to teach the corresponding knowledge in Chinese based on the user's questions: \textit{\textbf{[background information]}}.  Here is the information about the Scene around you: \textit{\textbf{[scene description]}}. Here is the message of the player: \textit{\textbf{[transcribed user voice input]}}. \\  
\bottomrule
\end{tabular}
\end{table}

\clearpage
\textbf{E: Prompts for Background Information}
\begin{table}[!htb]
\begin{tabular}{l p{12cm}}
\toprule
\textbf{Background Information}              & The Pavilion of Prince Teng, originally built during the Tang Dynasty by the order of Emperor Taizong of Tang, is located in Nanchang, Jiangxi Province. This pavilion is an outstanding example of ancient Chinese architecture and a confluence of cultural and political symbolism. From a modern perspective, the Pavilion of Prince Teng has been rebuilt 29 times throughout its history, with the latest reconstruction completed in 1989, demonstrating ongoing efforts to restore and preserve this cultural heritage. 

\vspace{1em}

Architecturally, the Pavilion of Prince Teng is renowned for its symmetrical design and the "three visible, seven hidden" layout, showcasing the exquisite craftsmanship and artistic aesthetics of Tang Dynasty architecture. The pavilion has seven floors, reaching a total height of 57.5 meters, with each floor intricately decorated, featuring elaborate wood carvings and paintings that reflect the artistic style of the Tang Dynasty. The Pavilion of Prince Teng is an awe-inspiring work of art both visually and culturally. On the sixth floor, a golden plaque inscribed with "Pavilion of Prince Teng" is the calligraphy of Su Shi. 

\vspace{1em}

In literature, the Pavilion of Prince Teng is immortalized by Wang Bo's "Preface to the Pavilion of Prince Teng." This piece is celebrated for its refined parallel prose style, complex literary techniques, such as antithesis and parallelism, and vivid metaphors depicting the natural scenery. The famous line "The sunset and the solitary duck fly together" from the preface greatly enriches the treasury of Chinese literature. Wang Bo's preface not only showcases his literary talent but also elevates the cultural status of the Pavilion of Prince Teng in history. Wang Bo, along with Yang Jiong, Lu Zhaolin, and Luo Binwang, is known as one of the "Four Talents of the Early Tang Dynasty." 

\vspace{1em}

Historically, the Pavilion of Prince Teng symbolized the display of royal power and cultural pride of the Jiangnan region in ancient times, representing a confluence of political authority and cultural prosperity. 

\vspace{1em}

Modern preservation efforts face numerous challenges, primarily maintaining the authenticity and integrity of this cultural heritage and addressing the environmental pressures from increasing numbers of visitors. The ancients referred to the Pavilion of Prince Teng as the "Water Pen," reflecting its role as the core of the city's feng shui. 

\vspace{1em}

Despite these challenges, the Pavilion of Prince Teng continues to attract visitors and scholars from around the world, serving as an important venue for studying Tang Dynasty architecture and literature. It is not only a historic building but also a symbol of Chinese culture, consistently showcasing its profound historical and cultural significance, embodying the union of political power and cultural flourishing. \\  
\bottomrule
\end{tabular}
\end{table}

\twocolumn


\end{document}